%% file: RJwrapper.tex
\documentclass[a4paper]{report}
\usepackage[utf8]{inputenc}
\usepackage[T1]{fontenc}
\usepackage{RJournal}
\usepackage{rotating}
\usepackage{amsmath,amssymb,array}
\usepackage{booktabs}

\usepackage{amsthm}

\newtheorem{Proposition}{Proposition}[section]
\newtheorem{Corollary}{Corollary}[section]

\begin{document}

\sectionhead{Contributed research article}
\volume{XX}
\volnumber{YY}
\year{20ZZ}
\month{AAAA}

\begin{article}
  \input{algaba-prieto-saavedra.tex}

\end{article}

\end{document}

%% file: algaba-prieto-saavedra.tex
\title{Rankings in the Zerkani network by a game theoretical approach}
\author{by Encarnación Algaba, Andrea Prieto and Alejandro Saavedra-Nieves}

\maketitle

\abstract{
This paper introduces the Banzhaf and Banzhaf-Owen values as novel centrality measures for ranking terrorists in a network. This new approach let integrate the complete topology (i.e. nodes and edges) of the network and a coalitional structure on the nodes of the network. More precisely, the characteristics of the nodes (e.g., terrorists) of the network and their possible relationships (e.g., types of communication links), as well as coalitional information (e.g. level of hierarchies) independent of the network.

First, for both centrality measures, we provide approximation algorithms and the corresponding R-codes. Second, as illustration, we rank the members of the Zerkani network, responsible for the attacks in Paris (2015) and Brussels (2016). Finally, we give a comparison between the rankings established by Banzhaf and Banzhaf-Owen and the rankings obtained when using the Shapley value (cf. \citealp{Hamers:2019}) and the Owen value as centrality measures.

}

\section{Introduction}

Over recent years, new jihadist cells have formed relatively easily in the Western World with the aim of committing terrorist acts. 
One of the recruitment organizations most active and dangerous is known as the Zerkani network. 

The Zerkani network was named after Khalid Zerkani, a Moroccan who was living in the Brussels municipality of Molenbeek and that was introduced into the network by Reda Kriket. Justice was able to prove that Zerkani indoctrinated very young people up to the point that they were willing to sacrifice themselves and to commit crimes to cover the worth for their journey to death. On November 13, 2015 the Islamic State carried out simultaneous attacks in several places in France, such as the Bataclan concert hall, killing more than a hundred people and wounding hundreds. Few months later, on March 22, 2016, part of those who were involved behind of the Paris attacks managed to launch another massive attack in Brussels, detonating suicide bombs at Zaventem International Airport and in Maelbeek subway station, killing thirty-two people and injuring many more. The main figures responsible for the tactical operations of the Paris and Brussels' attacks, that caused the death of more than one hundred seventy people and more than five hundred injured, were Abdelhamid Abaaoud and jihadist recruiter Khalid Zerkani. The Zerkani network provided personnel, training, planning, attack, escape, and evasion. Although the individual exact role of Zerkani is not entirely clear, he was found guilty of being one of those responsible for perpetrating both attacks mentioned above. 

However, they were backed by a network of people who supported them in their operations. In this sense, the members of this network seek to work together to cause as much damage as possible. The government and intelligence services collect information on terrorist attacks suffered for years, who and how carry out them and other possible threats. More concretely, they know information about individuals, their means of financing, communication or relationship between them. Due to the increase in both frequency and intensity of these attacks in recent years, and their damage to society, it is essential to enhance investigations to neutralise potential attack attempts. For this purpose, one of the key aspects is to identify and, immediately thereafter, determine the importance of each of the key members of the network. However, since the resources of the relevant agencies (for instance, the police or the national intelligence services, among others) are often limited for a variety of reasons, they must be allocated in an optimal way in order to stop potential terrorist attacks before they occur. Among many other things, a key issue is to identify the main leaders of a network.  It is well known that breaking up such a network, by identifying its members, immediately neutralises its criminal activity. This non-imminent risk allows the security forces to continue their investigations in a non-precipitous manner, which in turn maximises the effectiveness of their operations.

From a purely mathematical point of view, the data of a terrorist network can be studied through the construction of a graph. The terrorists become the nodes and the edges represent the interaction between each pair of these individuals. Once the network structure is built, there are several ways to rank their members by their relevance in the organization. The most used methods for ranking are based on standard network measures, such as degree centrality, betweenness, centrality and closeness centrality \citet{Koschade:2006}. However, the main drawbacks of this kind of measures focus on the fact of only considering the structure of the network. In this line of research, another approaches have been considered in literature as alternative. For instance, the standard social network centrality approach has been also used in analysing criminal and terror networks. See, for instance, \citet{Sparrow:1991}, \citet{Peterson:1994}, \citet{Klerks:2001}, \citet{Carley:2003}, or \citet{Farley:2003}, among others. 

One common issue to all these proposals is the fact that not considering valuable information about the communication between the members of the network. \citet{LindelaufHamers:2013}, \citet{Husslage:2015} solve these deficiencies, taking into account the heterogeneity of links and nodes through the use of cooperative game theory, which allows for adding this information by considering transferable utility games (or TU-games). Usually, TU-games are considered for the modelling of those multiagent interactive situations, in which collaboration of involved individuals in groups is a key aspect to achieve a common goal.  One of the  goals of TU-games consists of distributing the profits/costs that result from their cooperation. For this purpose, specific mechanisms as the coalitional values are introduced in the game-theoretic literature to determine payoff vectors. When applying them for the case of terrorist networks, a ranking of the members of the network according to its relevance can be determined. For instance, \citet{Hamers:2019} rank the members of the Zerkani network accordingly to the Shapley value \citep{Shapley:1953}for the TU-games of \cite{Husslage:2015}. However, other coalitional values, as the Banzhaf value \citep{Banzhaf:1964} for TU-games, can be used as a criterion. Although it is originally considered for voting situations, see \citep{Banzhaf:1964}, \cite{owen1975multilinear} generalizes its usage in other settings.  \cite{owen1986values} use it when restrictions of communication are represented by graphs, being characterized by \cite{alonso2006banzhaf}, in this scenario. Applications of the Banzhaf value also arise in electrical engineering \citep{chow1961characterization}, in computation \citep{ben1985collective}, in genetic \citep{lucchetti2010shapley} and even, it can be used as a design tool in coalitional control \citep{MUROS201721}. The use of the Shapley value and the Banzhaf value, that at first assume that all coalitions can be formed, can certainly be limited because there exist some situations in real-world in which cooperation among players may be restricted. Namely, transferable utility games with a priori unions (or TU-games with a priori unions) are introduced to model these situations, with multiple applications in fields as political science, logistics, or cost allocation problems. The  Owen value \citep{Owen:1977} extends the Shapley value for this class of games, and the Banzhaf-Owen value \citep{Owen:1982} do the same with the Banzhaf value for those situations with a priori unions.

In practice, the main drawbacks of these values are computational (see \cite{deng1994complexity}, for details). However, such solutions can be obtained in polynomial time by using alternative expressions for classical applications of cooperative game theory. See \cite{littlechild1973simple} and \cite{vazquez1997owen} for the cases of the Shapley value and the Owen value in airport games, respectively; \cite{fragnelli2000share} and \cite{costa2016polynomial} do the same for problems of maintenance of railways infrastructures; \cite{lucchetti2010shapley} provides the exact expression the Banzhaf value for micro-array games; or \cite{leech2002voting}, 
\cite{algaba2003computing}, \cite{alonso2005generating} or \cite{algaba2007distribution}, exactly compute  coalitional values in voting situations. For a more general setting, sampling methodologies \citep{cochran2007sampling} have become increasingly important as an alternative solution to the computational issues raised above. This is due to the fact that coalitional values are, in practice, seen as the population mean of marginal contributions under different probability mass functions. Thus, its analysis from a statistical point of view is natural, since their approximations are based on the average of a random subset of marginal vectors. We refer to \citet{Castro:2009} for the Shapley value estimation and to \citet{bachrach2010approximating} for the Banzhaf value estimation. In those settings with a priori unions, \citet{Saavedra:2018} and \cite{Saavedra:2020} describe sampling methodologies for estimation of the Owen value and the Banzhaf-Owen value, respectively. As immediate applications, \citet{SAAVEDRANIEVES2020655} estimate the random arrival rule (the Shapley value) for bankruptcy games, or  \citet{Hamers:2019} rank the terrorists belonging to a terrorist network by its relevance, using the Shapley value.

In this paper, we focus on ranking the members of the Zerkani network considering the existence of different degrees of relationships among them. To this aim, we use those coalitional values inspired by the approach of \citeauthor{Banzhaf:1964}. Firstly, we will consider the Banzhaf value, following the results obtained by \citet{Hamers:2019} for the Shapley value. In other direction, the possible affinities among the members of a network can be naturally described in terms of an a priori coalitional structure, which makes possible the extension through the definition of TU-games with a priori unions. This strongly justifies the usage of coalitional values, as the Owen value (see \citealp{Algaba:2021ch}) or the Banzhaf-Owen value, as mechanisms of ranking since let enrich the information used for obtaining the rankings. Due to the difficulties in computing, in an exact way, both of them, we make use of sampling techniques for approximating these results. Specifically, we consider the procedure in \cite{bachrach2010approximating} for estimating the Banzhaf value, and we slightly vary the proposal of \cite{Saavedra:2020}, by innovatively adding the hypothesis of replacement in sampling, for approximating the Banzhaf-Owen value. Once they are obtained, it is possible to rank the terrorists from the Zerkani network according to the decreasing order of both estimations.

The work is organised as follows. Section \ref{section1} introduces the basic notation on TU-games and on coalitional values in those settings with a priori unions. Section \ref{sampling:section} introduces the sampling methodologies for estimating the Banzhaf value and the Banzhaf-Owen value as new centrality measures. The data set used in this work is presented in Section \ref{ZERKANIanalysis}. Section \ref{compt:analysis} focuses on ranking the members of Zerkani network through the approximation of the Banzhaf value and the Banzhaf-Owen value by using R software. Section \ref{discussion} presents a brief comparison of the obtained rankings with respect to those obtained for the Shapley value and the Owen value, respectively.  Two appendices are also included. Appendix A numerically details the ranking obtained under the estimations of the Banzhaf value. Appendix B depicts the one obtained under the Banzhaf-Owen value estimation.

\section{Centrality measures for networks based on solutions for TU-games}\label{section1}

In this paper, we focus on cooperative games with transferable utility, called TU-games, where agents involved, called players, can work together and generate extra profits in comparison with the situation of doing so individually. The central issue of this class of games is how to adequately allocate the total profit among all the members of the game.

Formally, a transferable utility cooperative game is a pair $(N,v)$, where $N=\{1,2,...,n\}$ is the set of players, called the grand coalition, and $v: 2^{n} \longrightarrow \mathbb{R}$ is a map satisfying that $v(\emptyset)=0$. A coalition is a subset of players $S \subseteq N$ and $v(S)$ denotes the maximum value that players in $S$ can receive by the cooperation of its members. We denote by $\mathcal{G}^{N}$ the set of all TU-games with set of players $N$.

Here, we mainly follow the idea of \citet{LindelaufHamers:2013}, through the usage of two specific classes of games, which take into account both the structure of the network and the relational and the individual strength of the members of a given network. These two games have already been used to analyze the Zerkani network, although the first of them was slightly modified with respect to the original TU-game introduced in \citet{Husslage:2015}.

Formally, a network can be represented by an undirected graph $G=(N,E)$, where:
\begin{itemize}
    \item $N$ denotes the node set of the graph that represents the set of members of the network, and 
    \item $E$ is the set of links, that describes all relationships that exist between these members. A relationship between member $i$ and $j$ is denoted by $ij$, with $ij \in E$.
\end{itemize}   

Thus, if a coalition $S \subseteq N$ forms, the subnetwork $G_S$ is defined by the members of $S$ and its links in $E$, i.e., $G_S =(S,E_S)$ where \[E_S =\{ij \in E: i,j \in S\}.\]
 A coalition $S \subseteq N$ is said to be a connected coalition, if the network $G_S$ is connected, otherwise, $S$ is called disconnected.

Associated to any network $G=(N,E)$, the influence and the relations of individuals in a network $G$ are modelled through the following parameters:
\begin{itemize}
    \item The influence of  individuals in $G=(N,E)$ is  represented by a set of weights on player set $N$, i.e., ${\cal{I}}=\left \{ w_i \right \}_{i \in N}$ with $w_i \geq 0$.
    \item The relational strength between members of the network is given by a set of weights on the edges $E$, i.e., ${\cal{R}}= \{ k_{ij}\}_{ij \in E} $ with $k_{ij} \geq 0$.
\end{itemize}

\citet{Husslage:2015} introduced the \textit{ monotonic weighted connectivity TU-game (\text{or briefly, }mwconn)} $(N,v^{mwconn})$ with respect to the network $G=(N,E)$ based on ${\cal{I}}$ and ${\cal{R}}$ in the following way. For any connected coalition $S$, the characteristic function of $(N,v^{mwconn})$ is given by
\begin{equation}\label{e:mwconn}
v^{\text{mwconn}}(S) = f(S,{\cal{I},{\cal{R}}}),
\end{equation}being $f$ a context specific and tailor-made non-negative function depending on $S$, ${\cal{I}}$ and ${\cal{R}}$, which measures the effectiveness of coalitions in the network. In practice, it should be chosen to best reflect the situation and information at hand. In our setting, the function $f$ is specifically defined by
\begin{equation}\label{e:f}
f(S,\mathcal{I},\mathcal{R}) = 
\begin{cases}
\left( \sum_{j \in S} w_{j} \right) \cdot \mbox{max}_{lh\in E_{S}} k_{lh}, & \mbox{ if } |S|>1, \\
w_{S}, & \mbox{ if } |S|=1.
\end{cases}
\end{equation}

For a coalition $S$ that is disconnected, the characteristic function is defined by

\begin{equation}\label{e:mwnotconn}
v^{\text{mwconn}}(S) = \max_{T \in \Sigma_S} v^{\text{mwconn}}(T),
\end{equation}

\noindent
with $\Sigma_S$ the set of components (maximal connected coalitions) in $G_S$.
Observe, that the value of each disconnected coalition is based on the most effective component of this coalition. As it was mentioned, the main difference with the TU-games in \citet{LindelaufHamers:2013} and \citet{Husslage:2015} is that individuals can have a positive value now.

An \textit{ additive weighted connectivity TU-game (awconn)} $(N,v^{awconn})$ with respect to network $G=(N,E)$ based on ${\cal{I}}$ and ${\cal{R}}$ is, for a connected coalition $S$, defined by (\ref{e:mwconn}), identically to the corresponding monotonic weighted connectivity TU-game $(N,v^{mwconn})$. The difference between these two games is the definition of the disconnected coalitions. For a disconnected coalition $S$, the worth of $S$ in an additive weighted connectivity TU-game is given by
\begin{equation}\label{e:aconn}
v^{\text{awconn}}(S) = \sum_{T \in \Sigma_S} v^{\text{mwconn}}(T).
\end{equation}
 in such a way that all components or maximal connected subsets of $S$ are effective. This definition of game is consistent and it has been widely used to analyze networks derived from graphs as in communication situations in \cite{Myerson:1977} or in more general settings reflecting communication properties as hypergraphs communication situations \citep{vandenNouweland:1992}, union stable systems \citep{AlgabaMyerson:2001,AlgabaBilbao:2001}, or voting structures to describe problems in which there exists a feedback between the economic influence and the political power \citep{AlgabaE:2019}. Another framework is to analyze situations with hierarchical and communication features, simultaneously, as in \cite{AlgabaE:2018}.

A main issue in cooperative game theory focuses on distributing the total profit among all the members of $N$.
Therefore,  the definition of a payoff vector $x=(x_{i})_{i \in N} \in \mathbb{R}^{n}$ is required, being $x_{i}$ the payoff associated to player $i$ by its collaboration in a given TU-game $(N,v)$. In general, solution concepts (or simply, solutions) are introduced as maps $\phi : \mathcal{G}^{N} \to \mathbb{R}^{n}$ that assigns to each $(N,v)$ a set of payoff vectors. Attending to its cardinality (a point-valued or a set valued solution), solutions can be classified. Along this paper, we only focus in the case of those solutions providing a single point.

One of the most important solutions concepts is the Shapley value for a TU-game $(N,v)$, see \citet{Shapley:1953}. Take  $(N,v)\in \mathcal{G}^N$. The Shapley value for $(N,v)$ is formally defined as
\begin{equation}\label{e:shapmarg}
Sh_i(N,v)=\frac{1}{n!}\sum_{\pi\in \Pi} m_v^\pi(i), \text{ for all $i \in N$ and for every $(N,v)\in \mathcal{G}^{N}$,}
\end{equation}
\noindent where $m_v^\pi(i)$ denotes the marginal contribution of player $i$ in the order $\pi$, which is defined as $$m_v^\pi(i)=v\left(\{\pi_1,\ldots,\pi_k\}\right)-v\left(\{\pi_1,\ldots,\pi_{k-1}\}\right),$$ being $\Pi(N)$ the set of all possible orderings of the players in $N$. Among others, \citet{algaba2019handbook} is a comprehensive update of this value by illustrating its versatility of usage in multiple fields.

As an extension, those situations in which there exists an a priori coalitional structure that conditions their cooperative possibilities can be modelled also under this approach. The following concepts are required. A TU-game with a priori unions is given by a triplet $(N,v,P)$ where $(N,v) \in \mathcal{G}^{N}$ and $P=\{P_{1},...,P_{m}\}$ is a partition of $N$. In this case, we assume that $P$ is interpreted as the coalition structure that restricts the cooperation among the players in $N$.  $\mathcal{U}^{N}$ denotes the set of all TU-games with a priori unions with set of players $N$.

An important solution concept is the Owen value \citep{Owen:1977}, that extends the Shapley value to this setting.  Let $(N,v,P)$ be a TU-game with a priori unions. The Owen value (cf. \citealp{Owen:1977}) is an allocation rule that assigns, to each $i \in N$,
\begin{equation}\label{e:owen}
O_{i}(N,v,P) = \dfrac{1}{\left| \Pi_{P}(N)\right| } \displaystyle \sum_{\pi \in \Pi_{P}(N)} \left( v \left( P_{i}^{\pi}  \cup \left\lbrace i \right\rbrace \right) - v \left( P_{i}^{\pi} \right) \right),
\end{equation}
for every $(N,v,P)\in \mathcal{U}^{N}$, where $\Pi_{P}(N)$ denotes the set of all permutations of $N$ which are compatible with a coalition structure $P$. This means that the elements of each union of $P$ are not separated by $\pi$.Whereas, $P_{i}^{\pi}$ is the set of players that precede player $i$ in the order $\pi$. Earnestly, $\pi \in \Pi_{P}(N)$, if and only if, for all $i,j,k \in N$ it holds that
\[\mbox{if } i,j \in P_{i} \mbox{ and } \pi_{i}<\pi_{k} < \pi_{j} \Rightarrow k \in P_{i}.\]

Another well-known solution for TU-games, usually considered for simple games (although its usage is not restricted to general TU-games), is the Banzhaf value. The Banzhaf value \citep{Banzhaf:1964} is formally defined as
\begin{equation}\label{e:banzhaf}
Bz_{i}(N,v) = \dfrac{1}{2^{n-1}} \sum_{S \subseteq N \setminus \{i \}} (v(S \cup \{i\}) - v(S)), \text{ for all $i \in N$ and for every $(N,v)\in \mathcal{G}^{N}$.}
\end{equation}
Just like the Shapley value, this solution can be also interpreted as a weighted average of marginal contributions of a player to those coalitions that no contain it. The main difference lies on the corresponding weights used to average these addends. Despite not satisfying the efficiency property, i.e. $\sum_{i\in N}Bz_i(N,v)\neq v(N)$, the average of the marginal contributions of each player may be indicative of its influence on the overall set of players, as is the case in voting situations.

Besides, the Banzhaf-Owen value \citep{Owen:1982} is considered as the analogous extension of the Banzhaf value for TU-games with a priori unions. Take $(N,v,P) \in U^{N}$. This coalitional value assigns 
\begin{equation}\label{e:banzhafowen}
BzO_{i}(N,v,P)= \sum_{R \subseteq P \setminus P_{i}} \dfrac{1}{2^{m-1}} \sum_{S \subseteq P_{i} \setminus \{i\}} \dfrac{1}{2^{p_{i}-}1   } (v(\cup_{P_{l}\in R} P_{l} \cup S \cup \{i\}   )- v(\cup_{P_{l}\in R} P_{l} \cup S  ))
\end{equation}
for all $i \in N$ and for every $(N,v,P)\in \mathcal{U}^{N}$, where $P_{i} \in P$ such that $i \in P_{i}$ and $p_{i}= |P_{i}|$. Besides, a coalition $T \subseteq N \setminus \{i\}$ is compatible with partition $P$ for a player $i\in N$, if $T= \cup_{P_{l}\in R} P_{l} \cup S$ for a coalition of unions $R \subseteq P \setminus P_{i}$ and a coalition of players $S \subseteq P_{i} \setminus \{i\}$. 

In general, solutions for TU-games with a priori unions assume that the players in an union act jointly, so only contributions of a player to the coalitions formed by a subset of full unions and the players in its own union are averaged. In other words, both values differ from the previous ones, fixed a player $i$, because they only consider the possibility of joining the coalition of full unions and coalition of players from its same union.
However, if the coalition structure is formed by unitary unions or only by the grand coalition, the Banzhaf-Owen value and the Banzhaf value prescribe the same allocation. The same occurs with the Owen value and the Shapley value, respectively.

Once both approaches of TU-games for modelling networks are formally introduced, we note that, in line with \cite{Husslage:2015}, solution concepts can be applied for the two games for providing a game theoretic centrality measure. Let $G = (N, E)$ be a network  based on $\mathcal{I}$ and $\mathcal{R}$, $(N,v^{\text{mwconn}})$ and $(N,v^{\text{awconn}})$ the  TU-games associated to $G$ and take $i\in N$. Thus, game theoretic centrality measures $C^i$ for any $i\in N$ in $G$ can be directly given, for instance, by
\[C^i=Sh_i(N,v^{\text{mwconn}}) \mbox{ or }C^i=Sh_i(N,v^{\text{awconn}}),\]
if the Shapley value is considered as solution for the TU-games $(N,v^{\text{mwconn}})$ and $(N,v^{\text{awconn}})$, as in \citet{Hamers:2019}. Under the presence of a coalitional structure,  \cite{Algaba:2021ch} introduce the Owen value for both TU-games $(N,v^{\text{mwconn}})$ and $(N,v^{\text{awconn}})$ as new game theoretic centrality measures for networks. That is,  if $P$ is the associated partition of $N$,
\[C^i=O_i(N,v^{\text{mwconn}},P) \mbox{ or }C^i=O_i(N,v^{\text{awconn}},P).\]

Alternatively, if the Banzhaf solutions for TU-games are applied, the following game theoretical centrality measures innovatively arise. Firstly, if the Banzhaf value is considered, we can use 
\[C^i=Bz_i(N,v^{\text{mwconn}}) \mbox{ or }C^i=Bz_i(N,v^{\text{awconn}}).\]However, if the Banzhaf-Owen value is taken as solution under the presence of a coalitional structure, given by a partition $P$, we have that
\[C^i=BzO_i(N,v^{\text{mwconn}},P) \mbox{ or }C^i=BzO_i(N,v^{\text{awconn}},P).\]

In the remainder of the paper, we illustrate the usage of the Banzhaf value and the Banzhaf-Owen value with the only purpose of ranking the members of Zerkani network. Finally, we will compare the obtained results  with the ones given by the Shapley value and the Owen value.

\section{Sampling procedures to estimate the Banzhaf value and the Banzhaf-Owen value}\label{sampling:section}
Although the notion of marginal contribution of a player is intuitively clear, computing the Banzhaf value and the Banzhaf-Owen value becomes a hard task when the amount of players involved in the TU-game substantially increases. This fact justifies the needing of searching alternatives for providing good approximations of both solutions. 

Along this section, we formally describe some of the algorithms used for estimating the two above-mentioned values.

\subsection{A sampling procedure to estimate the Banzhaf value}\label{sectionEOwen}
We want to estimate the Banzhaf value of a TU-game  $(N,v)$. Following the ideas of \cite{bachrach2010approximating}, we formalize a procedure for estimating the Banzhaf value when the number of players involved is sufficiently large. The steps of the sampling procedure are the ones depicted below:

\begin{itemize}
	\item The population of the sampling procedure is the set of all coalitions of $N\setminus \{i\}$.
	\item The parameter to be estimated is $Bz_i(N,v)$, i.e. the player $i$'s Banzhaf value.
	\item The characteristic to study in each sampling unit, $T\subseteq N\setminus \{i\}$, is the player $i$'s marginal {contribution} to $T$. That is, \[x(T)_i=v(T\cup \{i\})-v(T).\]
	\item We take with replacement a sample of $\ell$  coalitions in $N\setminus \{i\}$, i.e.  $\mathcal{T}=\{T_1,\dots,T_{\ell}\}$, with $T_j\subseteq N\setminus \{i\}$ for all $j=1,\dots,\ell$ and $1< \ell \leq 2^{n-1}$.

	\item The estimation of $Bz_i(N,v)$, for every $i\in N$, is obtained as the mean of the 	marginal contributions over the sample, that is,  ${\overline{Bz}}_i=\frac{1}{\ell}\overset{\ell}{\underset{j=1}{\sum}}x(T_j)_i$ where $\ell$ denotes the \mbox{sampling} size.
\end{itemize}

Once we apply this procedure for each player, the vector ${\overline{Bz}}=({\overline{Bz}}_1,\dots,{\overline{Bz}}_n)$ {corres\-ponds} to the estimation of the Banzhaf value of all players in $(N,v)$. A fundamental issue in the problem focuses in bounding the  error in the estimation that is often not possible to be measured in practice. For this reason, the following probabilistic bound can be theoretically provided instead:
\begin{equation*}
\mathbb{P}(|{\overline{Bz}}_i-{Bz}_i|\geq \varepsilon)\leq\alpha,\mbox{ with }\varepsilon>0 \mbox{ and }\alpha \in (0,1].
\end{equation*}
Take $\varepsilon>0$, $\alpha\in (0,1)$  and let $(N,v)$ be a TU-game. Then, if we take	$r_i=\underset{R,R'\subseteq N\setminus \{i\}}{\max}(x(R)_i-x(R')_i)$, it satisfies that
\begin{equation}\label{hoeff}\ell\geq \min\bigg\{\frac{1}{4\alpha\varepsilon^2},\frac{\ln(2/\alpha)}{2\varepsilon^2}\bigg\}r_i^2\mbox{ implies that } \mathbb{P}(|{\overline{Bz}}_i-{Bz}_i|\geq \varepsilon)\leq \alpha.\end{equation}Thus, the estimated Banzhaf value usually becomes a good approximation of the real one when sampling sizes sufficiently enlarge. Those details focused on the bound based on Hoeffding's inequality can be found in \cite{bachrach2010approximating}.

Thus, determining the range of the marginal contributions in this setting plays a fundamental role in the analysis of the error. In what follows, we focus our attention in this task and we provide some specific results to bound the error in estimating centrality measures in networks such as those considered here.

\begin{Proposition}\label{ref_prop}
Let $(N,v^{\text{mwconn}}) \mbox{ or } (N,v^{\text{awconn}})$ be the monotonic or additive TU-games associated to a given network $G$. Thus, it satisfies, for every $i\in N$, that
\begin{equation}\label{rangeO}r_i=\left(\sum_{j\in N} w_{j} \right) \cdot \mbox{max}_{lh\in E_{N}} k_{lh}.\end{equation}
\end{Proposition}

\begin{proof}The scheme of the proof is divided into two parts, by distinguishing the monotonic and the additive approaches of the TU-games under consideration.

Firstly, using Equation (\ref{e:mwnotconn}), if $i\in N$ is an arbitrary player,  $S\subseteq N\setminus \{i\}$ is a non-connected coalition,  and $T^{\ast}={\arg\max}_{T \in \Sigma_{S\cup \{i\}}}v^{\text{mwconn}}(T)$, it easily follows that
\begin{equation} \label{eq1}
\begin{split}
v^{\text{mwconn}}(S\cup \{i\})-v^{\text{mwconn}}(S) & \leq v^{\text{mwconn}}(S\cup \{i\}) \\
& = \max_{T \in \Sigma_{S\cup \{i\}}} v^{\text{mwconn}}(T)\\
&=v^{\text{mwconn}}(T^{\ast})\\
& = \left( \sum_{j \in T^{\ast}} w_{j} \right) \cdot \mbox{max}_{lh\in E_{T^{\ast}}} k_{lh}\\
& \leq \left( \sum_{j \in N} w_{j} \right) \cdot \mbox{max}_{lh\in E_{N}} k_{lh},
\end{split}
\end{equation}
since that $\mbox{max}_{lh\in E_{T}} k_{lh}\leq \mbox{max}_{lh\in E_{N}} k_{lh}$, for any $T\subseteq N$.

Secondly, for the additive case, we get the same conclusions. Indeed, using Equation (\ref{e:aconn}), if we take $i\in N$ an arbitrary player and $S\subseteq N\setminus \{i\}$ a non-connected coalition, it holds that
\begin{equation} \label{eq2}
\begin{split}
v^{\text{awconn}}(S\cup \{i\})-v^{\text{awconn}}(S) & \leq v^{\text{awconn}}(S\cup \{i\}) \\
& = \sum_{T \in \Sigma_{S\cup\{i\}}} v^{\text{mwconn}}(T)\\
& =\sum_{T \in \Sigma_{S\cup\{i\}}}\left( \left( \sum_{j \in T} w_{j} \right) \cdot \mbox{max}_{lh\in E_{T}} k_{lh}\right)\\
& \leq\left(\sum_{T \in \Sigma_{S\cup\{i\}}} \left( \sum_{j \in T} w_{j} \right)\right) \cdot \mbox{max}_{lh\in E_{N}} k_{lh}\\
& \leq\left(\sum_{j \in N} w_{j} \right) \cdot \mbox{max}_{lh\in E_{N}} k_{lh},
\end{split}
\end{equation}
where, analogously, the second inequality is due to the fact of that $\mbox{max}_{lh\in E_{T}} k_{lh}\leq \mbox{max}_{lh\in E_{N}} k_{lh}$, for any $T\subseteq N$.

Hence, the marginal contributions always satisfies that $0\leq x(S)_i\leq \left(\sum_{j \in N} w_{j} \right) \cdot \mbox{max}_{lh\in E_{N}} k_{lh}$ for a fixed agent $i$ and for any coalition $S$ in $N$, in such way that it can be naturally tolerated that
\begin{equation*}r_i=\left(\sum_{j\in N} w_{j} \right) \cdot \mbox{max}_{lh\in E_{N}} k_{lh},\end{equation*}for every $i\in N$. This concludes the proof.
\end{proof}

The following corollary can be immediately obtained.

\begin{Corollary}\label{boundBZ}
Take $\varepsilon>0$, $\alpha\in (0,1)$  and $(N,v^{\text{mwconn}}) \mbox{ or } (N,v^{\text{awconn}})$ the monotonic or additive TU-games associated to a given network $G$. Thus, if we take  $\ell$ such that
\begin{equation}\label{hoeff_graph}\ell\geq \min\bigg\{\frac{1}{4\alpha\varepsilon^2},\frac{\ln(2/\alpha)}{2\varepsilon^2}\bigg\}\left(\left(\sum_{j \in N} w_{j} \right) \cdot \mbox{max}_{lh\in E_{N}} k_{lh}\right)^2,\end{equation}
then, $\mathbb{P}(|{\overline{Bz}}_i-{Bz}_i|\geq \varepsilon)\leq \alpha,$ for every $i\in N$.
\end{Corollary}

\subsection{A two-stage sampling procedure to estimate the Banzhaf-Owen value}\label{sectionEBOwen}

Now, we follow the ideas of the procedure in \cite{Saavedra:2020} to approximate the Banzhaf-Owen value. 

Given a TU-game with a priori coalitional structure $(N,v,P)$, with $P=\{P_{1},...,P_{m}\}$ and a fixed arbitrary player $i \in N$, the two-stage procedure for estimating the Banzhaf-Owen value is described below:
\begin{itemize}
	\item The population of the sampling process is the set of all compatible coalitions with $P$ for $i$. 
	\item The parameter to be estimated is  $BzO_{i}(N,v,P)$, for all $i \in N$.
	\item The characteristic to be studied in each sampling unit corresponds to player $i's$ marginal contribution for each coalition $T$ that is compatible with $P$ for $i$. Then, if we consider $T \subseteq N \setminus \{i\}$ in terms of the $R = \{ P_{k} : P_{k} \subset T  \}$ and $S = T \cap P_{(i)}$, we have 
	$$ x(R,S)_{i} = v(\cup_{P_{l}\in R} P_{l} \cup S \cup \{i\}) - v(\cup_{P_{l}\in R}  P_{l} \cup S).$$
	\item The sampling procedure involves two steps: 
	\begin{itemize}
		\item Firstly, we take with replacement a sample $\mathcal{R} = \{ R_{1},...,R_{\ell_{r}}  \}$ of $\ell_{r}$ coalitions $R_{j} \subseteq P \setminus P_{(i)}$.
		\item After, for every $R_{j} \in \mathcal{R}$, we choose with replacement a sample $\mathcal{S}_{R_{j}} = \{S_{j1},\dots,S_{j\ell_{s}}\}$ of $\ell_{s}$ coalitions $S_{jk} \subseteq P_{(i)} \setminus \{i\}$.
	\end{itemize}
	as a result, we obtain a sample of $\ell_{r}\ell_{s}$ compatible coalitions, where each element takes the form $\cup_{P_{l}\in R} P_{l} \cup S_{jk}$ for $j=1,\dots,\ell_{r}$, with $1 \leq \ell_{r} \leq 2^{m-1}$, and $k=1,\dots,\ell_{s}$, with $1 \leq l_{s} \leq 2^{p_{i}-1}$.
	\item The mean of the marginal contributions vectors over the sample corresponds to the estimation of the Banzhaf-Owen value. That is,
	\[{\overline{BzO}}_{i} = \frac{1}{\ell_{r}} \sum_{j=1}^{\ell_{r}}\left(\frac{1}{\ell_{s}} \sum_{k=1}^{\ell_{s}}x(R_{j},S_{jk})_{i}\right)\]	
	where $\ell_r$ and $\ell_s$ are the sampling sizes.	
\end{itemize}

By applying this procedure for all $i\in N$, ${\overline{BzO}}=({\overline{BzO}}_1,\dots,{\overline{BzO}}_n)$ is the estimation of the Banzhaf-Owen value for $(N,v,P)$. Regarding to the Statistical properties, we use its interpretation as a mean of means. Thus, we additionally define  $\overline{BzO}_i^{R_j}=\frac{1}{\ell_s}\overset{\ell_s}{\underset{k=1}{\sum}}x(R_j,S_{jk})_i$ as the unbiased estimator of $BzO_i^{R_j}=\frac{1}{2^{p_i-1}}\underset{S\subseteq P_{(i)}\setminus \{i\}}{\sum}x(R_j,S)$, that is, the theoretical mean of  player $i$'s marginal contributions  using all compatible coalitions associated to a coalition of unions $R_j$.

Below, we focus on analysing the properties of the estimator of the Banzhaf-Owen value of player $i$, ${\overline{BzO}}_i$, from a Statistical perspective. First, according to the definition of this estimator, it is unbiased because 
\begin{equation*}\mathbb{E}({\overline{BzO}}_i) =
\mathbb{E}_{1}(\mathbb{E}_{2}({\overline{BzO}}_i))={{BzO}}_i,
\end{equation*}
being $\mathbb{E}_1(\cdot)$ and $\mathbb{E}_2(\cdot)$ the mean operators in  coalitions of unions and in coalitions into the union to which $i$ belongs to, respectively.

Besides, if $\mbox{Var}_1(\cdot)$ and $\mbox{Var}_2(\cdot)$ are the operators for variances with respect to the coalitions of unions and with respect to the coalitions into the unions to which $i$ belongs, respectively, the variance of ${\overline{BzO}}_i$ is
\begin{equation}\label{varformula}\mbox{Var}({\overline{BzO}}_i)=\mbox{Var}_{1}(\mathbb{E}_{2}({\overline{BzO}}_i))+\mathbb{E}_{1}(\mbox{Var}_{2}({\overline{BzO}}_i))\end{equation}
or, equivalently, is given by
\begin{align}\label{variance}\mbox{Var}({\overline{BzO}}_i)&
=\frac{1}{\ell_r}\bigg({{\theta}_a^2}+\frac{{\theta}_b^2}{\ell_s}\bigg),\end{align}
where 
\[\theta_a^2=\frac{1}{2^{m-1}-1}\underset{R\subseteq P\setminus P_{(i)}}{{\sum}}\big({BzO}_i^R-{{BzO}}_i \big)^2,\mbox{ and }\]
\begin{equation*}\label{thetab}\theta_b^2=\frac{1}{2^{m-1}}\underset{R\subseteq P\setminus P_{(i)}}{{\sum}}\bigg(\frac{1}{2^{p_i-1}-1}\underset{S\subseteq P_{(i)}\setminus\{i\}}{{\sum}}\big(x(R,S)_i-{{BzO}}^R_i \big)^2\bigg).\end{equation*}
Roughly speaking, $\theta_a^2$ refers to the variability of the means of unions with respect to $BzO_i$, $\theta_b^2$ denotes the average of the variances of the marginal contributions with respect to its theoretical mean $BzO_i^R$ for each $R\subseteq P\setminus P_{(i)}$. 

Notice that this proposal is slightly different to one proposed in \cite{Saavedra:2020}, since that approach involved the hypothesis of non-replacement in sampling. However, the main properties on the estimator are still satisfied and the main conclusions about the difficulties in obtaining probabilistic bounds of the incurred error maintain too. 

Next, we numerically analyse the rankings of the individuals in the Zerkani network, by approximating the Banzhaf value and the Banzhaf-Owen value for the \emph{mwconn} and \emph{awconn} TU-games. Notice that both require of approximation methodologies in this setting since the Zerkani network contains 47 members.

\section{The Zerkani network analysis}\label{ZERKANIanalysis}

On November 13, 2015 the Islamic State carried out simultaneous attacks in several places in France, such as outside the Bataclan concert hall, killing more than a hundred people and wounding hundreds. Few months later, on March 22, 2016, part of those who were involved behind of the Paris attacks managed to launch another massive attack in Brussels, detonating suicide bombs at Zaventem International Airport and in Maelbeek subway station, killing thirty two people and injuring many more. The main figures responsible for the tactical operations of the Paris and Brussels' attacks were Abdelhamid Abaaoud and jihadist recruiter Khalid Zerkani. The Zerkani network provided personnel, training, planning, attack, escape and evasion.

The aim of this paper focuses on ranking of the members of the Zerkani network according to their individual roles by approximating two solutions for TU-games, the Banzhaf value and the Banzhaf-Owen value, for the \emph{mwconn} and the \emph{awconn} TU-games, in (\ref{e:mwnotconn}) and in (\ref{e:aconn}) respectively. It has been necessary to use the approximating procedures considered in Section \ref{sampling:section}, since the Zerkani network contains 47 individuals. We compare the rankings that we have obtained and show the analysis computational of the process.

\begin{figure}[h!]
	\centering
	\includegraphics[width=\textwidth, height=14cm]{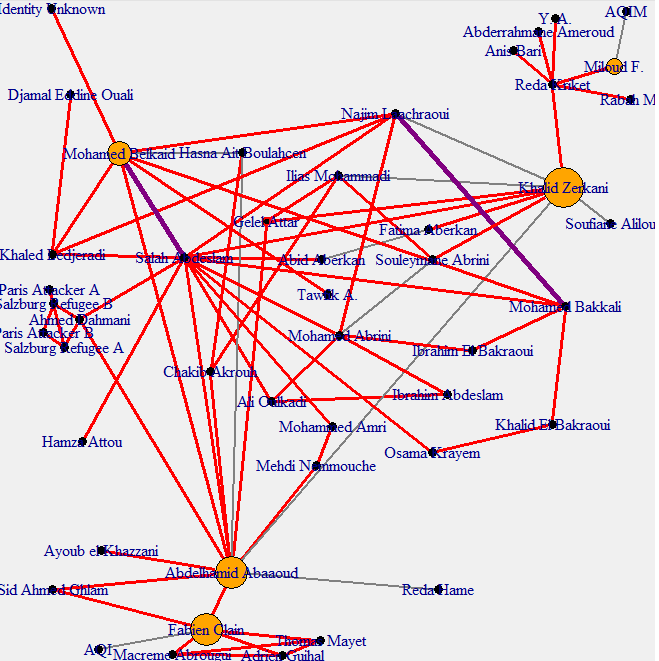}
	\caption{Graph of the Zerkani network.}
	\label{zerkaninetwork}
\end{figure}

Firstly, its associated graph is displayed in Figure \ref{zerkaninetwork}.

In our study we focus on the top-$10$ of the rankings, although we completely rank all 47 individuals. It is well-known that scarce resources limit intelligence and law enforcement agencies in their operations and, as a result, not all the time everybody terrorists involved may be kept under surveillance. For this reason, the most important task consists of identifying the most dangerous terrorists.

In order to achieve our aim of establishing rankings of terrorists, we consider two quite different schemes, although both of them are based on the Banzhaf value. Firstly, we studied the problem from a myopic perspective, by not considering the possible interactions between network members. After this, we take into account their possible affinities in terms of an adequate partition of the terrorists in the Zerkani network since adding this interesting information may be fundamental. The analysis of this kind of problem is still nascent (see \cite{Algaba:2021ch} for the case of the Owen value). The main novelty lies in the fact of that the Banzhaf-Owen value was never used to analyse a terrorist network. In those settings with a priori unions, as the information about the individuals are limited, several partitions have been study. It is important to note that this paper focuses on describing the performance of ranking procedures from a purely quantitative point of view. However, the task of choosing the appropriate partition is not trivial, as information about the relationships between terrorists, which is handled by security teams, is often not available to society as a whole. As the \emph{mwconn} and \emph{awconn} games already consider network connectivity, as well as the weights of individuals and their relationships, we have taken into account in this paper only the partition describing the features of the terrorists in the network.  More specifically, the considered partition has ten unions, $P=\{P_{1},P_2,\cdots,P_{10}\}$. The first union, $P_{1}$, groups the high ranks of the network, i.e. majors and those devoted to recruit terrorist as Abdelhamid Abaaoud, Fabien Clain or Khalid Zerkani (who have assigned a weight bigger than one). Union $P_{2}$ corresponds to the associated to the upper-level charges. Next, it is considered an inferior rank, i.e., those who have been recruited or who are under the authority of a major, leading to union $P_{3}$.  One of the relationships to take into account is to travel with, since during travels can be created strong relationships or it is assumed that can be discovered hide intentions, it gives way to unions $P_{4}$, $P_{5}$ y $P_{6}$. Moreover, inside the Zerkani network, there are several  terrorists who also belong to another network, network Kriket and they form another union, $P_{7}$. In the same way, there are also two individuals associated in $Forging$ $Ring$, grouped in $P_{8}$. The two last groups are due to that two individuals were arrested simultaneously in $Forest$ and they are associated without having more details about it, giving way to $P_{9}$ and $P_{10}$.

\section{Computational analysis of Zerkani network}\label{compt:analysis}
Below, we address into the analysis, from a purely computational point of view, of the problem of obtaining the rankings of the top ten terrorists within the Zerkani network, using the approximations of the Banzhaf value and of the Banzhaf-Owen value, respectively. In addition to the numerical results associated to these approximations, the R code required to their obtaining in both scenarios is described.

Along this section, the required objects and functions have been defined in R software. Then, the characteristic functions of the monotonic weighted connectivity TU-game $(mwconn)$ and the additive weighted connectivity TU-game $(awconn)$ are also implemented for the Banzhaf value and for the Banzhaf-Owen value estimation in the Zerkani network.

\subsection{Graph of the Zerkani network}
Firstly, the network is built by using the information of the 47 individuals belonging to it. This information has given rise to definition of 11 possible relationships. Since there are several individuals who are linked through two of these links, there are a total of 13 weights corresponding to the different connections between terrorists.

In order to deal with graphs in R software, the usage of the following packages are required. Notice that the fact of that \verb|igraph| package of R does not support the possibility of parallelising the code limits enough their computation. 
\begin{example}
library("igraph")	
library("dplyr")
\end{example}

The file \verb|is_Zerkani_network.csv| contains the available dataset about the members of the network that was collected by intelligence services. Due to the fact that the information we use is real, it can not be openly supplied. From this moment, \verb|ZerkaniNetwork| contains all the information extracted from our data base.

\begin{example}
ZerkaniNetwork=read.csv("is_Zerkani_network.csv",header=T,sep=",")
\end{example}

By means of the possible relations, different extra weights are assigned to the edges, since initially they all have value 1. The same happens with the members, also an amount is added to their nodes. Thus,  the final graph is obtained. First, we define a vector of characters that contain all possibilities of relationship among the members of Zerkani network as follows:

\begin{example}
Relationships=c("Associate of","Brother of","Commander of","Family relationship",
		"Funded","Lived with","Nephew of","Recruiter of","Supporter of",
		"Traveled to Syria with","Traveled with", "Associate and traveled with",
		"Traveled and lived with")
\end{example}

Since there are two terrorists with more than one relationship with each other their linkage and dangerousness is higher. Thus, we built the following vectors of weights for the set of possible relationships.
\begin{example}
WeightR=c(2,1,2,1,1,2,1,1,1,2,2,4,4)
WeightEx=c(0,0,2,0,2,0,0,1,1,0,0)
\end{example}

By considering these values, we can obtain a new vector with the weights associated to the edges of the resulting graph as follows:

\begin{example}
ZData=data.frame(ZerkaniNetwork$Entity.A,ZerkaniNetwork$Entity.B)
Zgraph=graph.data.frame(ZData,directed=FALSE)
WeightI=rep(0,length(ZerkaniNetwork$Entity.A))
for (kk in 1:length(ZerkaniNetwork$Entity.A)){
	WeightI[kk]=WeightR[which(ZerkaniNetwork$Relationship[kk]==Relationships)]+1
}
E(Zgraph)$width=WeightI
Zedges=data.frame(ZData,E(Zgraph)$width)
colnames(Zedges)=c("vi","vj","Weight")
\end{example}

Initially, each node has associated a weight equal to 1, but as we want to see who has a greater weight the following modification should be done:
\begin{example}
V(Zgraph)$size=rep(4,length(V(Zgraph)))
MembersEx=c("Abdelhamid Abaaoud","Fabien Clain","Khalid Zerkani","Miloud F.","Mohamed Belkaid")
WeightMembersEx=c(4,4,5,2,3)
VertexME=rep(0,length(WeightMembersEx))
for (kk in 1:length(WeightMembersEx)){
	VertexME[kk]=which(MembersEx[kk]==vertex.attributes(Zgraph,index =V(Zgraph))$name)
}
\end{example}
From a graphical approach, if we prefer viewing the nodes bigger, we have firstly  to do:
\begin{example}
V(Zgraph)[VertexME]$size=4.*WeightMembersEx 
Individuals=(V(Zgraph)$size)
\end{example}

%
%
%
%
%

For instance, to add colour to the graph, as well as, to increase the weight of the nodes  to be able to differentiate them better, we have to run the following lines of R code. Notice that  the correct data will be still used to internally work. 
\begin{example}	
V(Zgraph)[which(Individuals>1)]$color="orange"
V(Zgraph)[which(Individuals==1)]$color="black"
E(Zgraph)[which(WeightI==2)]$color="grey"
E(Zgraph)[which(WeightI==3)]$color="red"
E(Zgraph)[which(WeightI==5)]$color="purple"
\end{example}

To conclude, if we execute the following line in R software, we can obtain a graphical representation as the one depicted in Figure \ref{zerkaninetwork}.
\begin{example}	
plot(Zgraph)
	\end{example}

\subsection{Definition of the characteristic functions}
			Before using the sampling methods to estimate the Banzhaf value and the Banzhaf-Owen value, it is necessary to define the characteristic functions of the games to which they are applied. 
			
			Thus, the function of the monotonic TU-game $v^{\mbox{mwconn}}$ is given, for each $S\subseteq N$, by
			$$ v^{\mbox{mwconn}}(S) = \begin{cases} f(S,\mathcal{I},\mathcal{R}), & \mbox{if } S \medspace \mbox{ connected,} \\
			\mbox{max}_{T \in \sum_{S}} v^{\mbox{mwconn}}(T), & \mbox{if } S \medspace \mbox{ disconnected.}
			\end{cases}$$ 
			and the characteristic function of the additive game, $v^{\mbox{awconn}}$, is, for a given coalition $S\subseteq N$, 
			$$v^{\mbox{awconn}}(S) = 
			\begin{cases}
			v^{\mbox{mwconn}}(S), & \mbox{if } S \medspace \mbox{  connected,} \\
			\sum_{T \in \sum_{S}} v^{\mbox{mwconn}}(T), & \mbox{if } S \medspace \mbox{ disconnected.}
			\end{cases}$$
			In order to make the program more effective, we have joined both functions into one, that is named as \verb|mawconn| function. The R code associated to this function is displayed next.
			\begin{example}
mawconn=function(S,I,R){
	nC<-(count_components(S))
	
	if(nC==1){
		f_SIR=f(S,I,R)
		result=c(f_SIR,f_SIR)
	}else{
		accum=0
		max=0
		dS<-decompose(S)
		for(i in 1:nC){
			value=f(dS[[i]],I,R)
			accum=accum+value #for awconn the sum of all the values
			if(value>max){max=value} #for mwconn the maximum value
		}
		result=c(max,accum)
	}
	result  
}	
			\end{example} 
			In this case, $f$ is defined so that it takes into account both individual strength and that of relationships between individuals. 
			$$f(S,\mathcal{I},\mathcal{R}) = \begin{cases} \left( \sum_{i \in S} w_{i}  \right) \cdot \mbox{max}_{lh \in E_{S}} k_{lh}, & \mbox{if } |S|>1, \\
			w_{S}, & \mbox{if } |S|=1.
			\end{cases}$$
			
We also implement this function in R software as follows.

\begin{example}
f=function(S,I,R){
	maxim=0
	if ((length(components(S)$membership))!=1){
		accum=0
		for(j in 1:length(V(S))){
			accum=accum+I[which(V(S)$name[j]==
			vertex.attributes(Zgraph,index = V(Zgraph))$name)]
		}
		geS<-get.edgelist(S)
		for(k in 1:dim(geS)[1]){
			v1=geS[k,1];2=geS[k,2]
			aux<-filter(R,vi == v1 & vj == v2 )$Weight
			if(length(aux)!=0){
				posmax=aux
			}else{
				posmax=(filter(R,vj == v1 & vi == v2 )$Weight)
			}
			if(posmax > maxim){maxim=posmax}
		}
		result=accum*maxim
	}else{
	    result=I[which(V(S)$name[1]==vertex.attributes(Zgraph,index = V(Zgraph))$name)]
	}
	result
}

	\end{example}

Notice that both functions have the same input arguments that are enumerated in the following list.
\begin{itemize}
\item \verb|S|: this argument indicates the coalition $S$ to be evaluated.
\item \verb|I|: this argument describes the set of weights on players set $N$, i.e., ${\cal{I}}=\left \{ w_i \right \}_{i \in N}$ with $w_i \geq 0$.
\item \verb|R|: this argument describes the set of weights on the the edges $E$, i.e., ${\cal{R}}=\left \{ k_{lh} \right \}_{lh \in E} $ with $k_{lh} \geq 0$.
\end{itemize}\verb|mawconn()| function returns a double output that are the numerical values of the characteristic function of the monotonic and additive games evaluated on $(S,\mathcal{I},\mathcal{R})$. Besides, \verb|f()| function returns the value of $f(S,\mathcal{I},\mathcal{R})$ for the indicated arguments.

\subsection{The Banzhaf value approximation in Zerkani network}
One of the objectives is to analyse the $top-10$ of the ranking according to the relevance of the terrorists given by the estimation of the Banzhaf value in the Zerkani network, in line with \citet{Hamers:2019}, who used an approximation of the Shapley value.  In addition, $(N,v^{mwconn})$ and $(N,v^{awconn})$ are not convex in general and therefore, it is not possible the usage of some properties that would reduce the computational complexity in its obtaining. This section describes the R code specifically built for  applying the procedure in Section \ref{sectionEOwen} to Zerkani network.

\subsubsection{R code}
Here, we show the R code used to rank the members of the Zerkani network accordingly to the non-increasing order of the components of the estimated Banzhaf value. Firstly, we describe the \verb|estimate_BZ()| function, that is, the R function built with the only purpose of estimating the Banzhaf value in the Zerkani network. This function has associated the following arguments as inputs:
\begin{itemize}
	\item \verb|seed|: a seed specifies the start point when a computer generates a random collection of permutations. That is, we can replicate the results when using the function in any computer and with the same value of the seed for a given set of parameters.
	\item \verb|nitmax|: a natural number which indicates the  number of samples used in the estimation.
		\item \verb|Individuals|: a vector that contains the individual weights of the terrorists involved in Zerkani network for measuring the influence of each of them.
	\item \verb|Zgraph|: a \verb|igraph| object that contains all the information associated to the graph induced by Zerkani network through the system of partitions that describe the possible affinities among its members.
	\item \verb|ZedgesNumber|: a \verb|data.frame| object that describes the edges of the graph in terms of nodes as well as its associated weight.
\end{itemize}

As output, this function returns a list of three elements that contains the following results:
\begin{itemize}
	\item An estimation of the Banzhaf value for the TU-game $(N,v^{\mbox{awconn}})$.
	\item An estimation of the Banzhaf value for the TU-game $(N,v^{\mbox{mwconn}})$.
	\item Information about the processing times (in seconds) in terms of the User time, System time and Elapsed time that R software provides by using the basic function of R \verb|proc.time()|.
\end{itemize}

The R code that contains the function required for the Banzhaf value estimation is displayed below.
\begin{example}
estimate_BZ<-function(seed,nitmax,Individuals,Zgraph,ZedgesNumber){

set.seed(seed)
n<-length(Individuals);dim<-n

BzA<-rep(0,n);BzW<-rep(0,n)

start<-proc.time()
for (i in 1:n){
	nitmax<-min(nitmax,2^(n-1))

	cont<-0
	ContributionW<-0
	ContributionA<-0

	while(cont<nitmax){
		Xn<-c()
		Xn<-sample(c(0,1),n,replace=T)
		Xn[i]=0
		im<-which(Xn>0)
		iml<-length(im)
		S=im
		Sui<-c(im,i)
		if (length(Sui)==1){
			valueindividual=Individuals[Sui]
			ContributionWS=ContributionWS+valueindividual
			ContributionAS=ContributionAS+valueindividual
		} else {
			Zgraphwithi=induced_subgraph(Zgraph,Sui)
			Zgraphwithouti=induced_subgraph(Zgraph,S)
                        valuewith=mawconn(Zgraphwithi,Individuals,ZedgesNumber)
                        valuewithout=mawconn(Zgraphwithouti,Individuals,ZedgesNumber)
                        ContributionWS=valuewith[1]-valuewithout[1]
                        ContributionAS=valuewith[2]-valuewithout[2]
		}
		ContributionW=ContributionW+ContributionWS
		ContributionA=ContributionA+ContributionAS
		cont=cont+1
	}
    BzA[i]<-ContributionA/cont
    BzW[i]<-ContributionW/cont
}

end<-proc.time()-start

return(list(BzA,BzW,as.vector(end)[1:3]))
}
\end{example}

As example of usage, the next code will provide an estimation of the Banzhaf value with $\ell=10^3$.
\begin{example}
BZ_Z<-estimate_BZ(1,10^3,Individuals,Zgraph,ZedgesNumber)
\end{example}
Note that this procedure is easily parallelisable, so that the existence of several processors considerably reduces the computational complexity needed for its determination, as well as the time required.

After the obtaining of the estimation of the Banzhaf value for each terrorist of Zerkani network under the both considered approaches, we can build the associated rankings of the most dangerous members given by the largest components of the Banzhaf value. Only the first 10 are stored. To this aim, we provide \verb|ranking()| function that considers as inputs:
\begin{itemize}
	\item \verb|allocation|: a numerical vector that contains the allocation vector used as criteria for ranking the terrorists. 
	\item \verb|names|: a vector with the labels that identify each terrorist. For instance, we can use their names.
\end{itemize}
This function returns a data frame with list of the terrorists that are ordered according to the non-increasing order of the allocation considered. The R code associated is displayed below.
\begin{example}
ranking<-function(allocation,names){
	dim<-length(allocation)
	orderplayers=sort(allocation,decreasing = TRUE)
	positions=rep(0,dim)
	i=1
	while(i<=dim){
		aux<-which(orderplayers[i]==allocation)
		if (length(aux)==1){
			positions[i]=which(orderplayers[i]==allocation)
			i=i+1
		} else {
			positions[i:(i+length(aux)-1)]<-which(orderplayers[i]==allocation)
			i=i+length(aux)
		}
	}
	Ranking=data.frame(names[positions],allocation[positions])
	names(Ranking)<-c("Terrorist","Allocation")
	return(Ranking)
}
\end{example}
In practice, the following lines provide the complete ranking of the terrorists  belonging to Zerkani network according to the estimated Banzhaf value.
\begin{example}
completenames<-V(Zgraph)$name
ranking(BZ_Z[[1]],completenames)
ranking(BZ_Z[[2]],completenames)
\end{example}

\subsubsection{Numerical results}

As we mentioned along the paper, one of the purpose is to rank the members of the Zerkani network according to the decreasing order of those allocations provided by the Banzhaf value. We use the R code described in this section.

By simplicity, we obtain $1000$ estimations of the Banzhaf value by using the sampling procedure described in Section \ref{sectionEOwen} with a sample size equal to $\ell=1000$ for the TU-games $(N,v^{mwconn})$ and $(N,v^{awconn})$, respectively. Besides, we use the theoretical properties that satisfy the resulting estimator to obtain a more exact estimation. By the Central Limit Theorem, when averaging all of these $1000$ approximations, the final result is equivalent to obtain a only estimation with $\ell=10^6$ in both cases. Table \ref{errors} shows the theoretical errors provided by Corollary (\ref{boundBZ}) for the problem of estimating the Banzhaf value when $r_i=300$, as Proposition (\ref{ref_prop}) ensures for the case of Zerkani network.

\begin{table}[h!]
	\begin{center}\resizebox{0.43\textwidth}{!}{
		\begin{tabular}{ p{1.25cm} p{1.4cm} p{1.4cm} p{1.4cm} }
			\hline
			&$\alpha=0.1$ &$\alpha=0.05$&$\alpha=0.01$\\			\hline \hline
			$\ell=10^3$ & 1.34069 &1.48773& 1.78297\\
			$\ell=10^6$  & 0.04240 &0.04705& 0.05638\\
			\hline 
		\end{tabular}}
		\caption{Theoretical errors ($\varepsilon$) for $\ell=10^3$ and $\ell=10^6$.}
		\label{errors}
	\end{center}
\end{table}
\vspace*{-0.5 cm}Table \ref{rank7_Banzhaf} depicts the Top-10 list of terrorists belonging to the Zerkani network and the corresponding results. More details can be found in Appendix A that enumerate the overall list of the members of Zerkani network. We remark the first 10 positions and we check some differences between the monotonic TU-game $(N,v^{mwconn})$ and the additive TU-game $(N,v^{awconn})$. See more details in Table \ref{rank7_Bz_complete}. Khalid Zerkani, who is considered to be the leader of the network, goes to the third position under both approaches.  Abdelhamid Abaaoud occupies the first position under the monotonic point of view and he moves to the fourth position under the second point of view. However, Mohamed Belkaid moves up from the fourth position in the monotonic scheme to the first position under additivity. Analogous comments can be extracted from the remainder of the list of members in the top 10, that at most of them are in the same positions.

\begin{table}[h!]
	\begin{center}\resizebox{0.9\textwidth}{!}{
		\begin{tabular}{ p{0.5cm}|p{4cm} p{2cm}|p{4cm} p{2cm}}
			\hline
			& \multicolumn{2}{c|}{Ranking $R_{mwconn}$} & \multicolumn{2}{c}{Ranking $R_{awconn}$}\\\hline
			Pos. & Terrorist & ${\overline{Bz}}$ & Terrorist & ${\overline{Bz}}$ \\
			\hline \hline
1	&	Abdelhamid Abaaoud	&	38.326053	&	Mohamed Belkaid	&	31.411917	\\
2	&	Salah Abdeslam	&	35.073561	&	Salah Abdeslam	&	26.167231	\\
3	&	Khalid Zerkani	&	33.930235	&	Khalid Zerkani	&	25.716073	\\
4	&	Mohamed Belkaid	&	33.267557	&	Abdelhamid Abaaoud	&	24.155854	\\
5	&	Najim Laachraoui	&	18.774721	&	Mohamed Bakkali	&	18.297970	\\
6	&	Mohamed Bakkali	&	18.307468	&	Najim Laachraoui	&	17.311352	\\
7	&	Fabien Clain	&	11.891287	&	Fabien Clain	&	16.512582	\\
8	&	Reda Kriket	&	8.316217	&	Reda Kriket	&	10.625338	\\
9	&	Ahmed Dahmani	&	8.129191	&	Mohamed Abrini	&	6.257625	\\
10	&	Mohamed Abrini	&	6.94882	&	Miloud F.	&	5.832491	\\
			\hline 
		\end{tabular}}
		\caption{Top-10 of the  ranking of terrorists in the Zerkani network, according to the estimated Banzhaf value for $(N,v^{mwconn})$ and $(N,v^{awconn})$ with $\ell=10^6$.}
		\label{rank7_Banzhaf}
	\end{center}
\end{table}

\vspace*{-0.5 cm}Once the Banzhaf value is estimated, we check how the sampling proposal for approximating the Banzhaf value performs in this example in terms of the variability of the results in this simulation study. By construction, we have obtained the results in Table \ref{rank7_Banzhaf} by averaging 1000 estimations of the Banzhaf value for the Zerkani network by using sample sizes equal to $\ell=10^3$. We separately focus our attention in those $10$ terrorists belonging to the top of the considered rankings.

\begin{table}[h!]
	\begin{center}\resizebox{0.9\textwidth}{!}{
	\begin{tabular}{ p{0.3cm}| p{4cm}| p{1cm} p{1.1cm} p{1cm}  p{0.9cm} p{1.15cm} p{1 cm}| p{0.6cm} }
			\hline
			& Terrorist&Min.&1st Qu.&Median&Mean&3rd Qu.&Max. & CV  \\			\hline \hline
1&Abdelhamid Abaaoud&36.741&37.990&38.323&38.326&38.693&40.209&0.014\\
2&Salah Abdeslam&32.228&34.490&35.092&35.074&35.637&37.687&0.024\\
3&Khalid Zerkani&32.044&33.527&33.904&33.9305&34.326&35.867&0.018\\
4&Mohamed Belkaid&30.890&32.744&33.260&33.268&33.790&35.804&0.024\\
5&Najim Laachraoui&16.250&18.347&18.794&18.775&19.239&20.773&0.037\\
6&Mohamed Bakkali&16.361&17.883&18.305&18.308&18.744&20.740&0.036\\
7&Fabien Clain&10.578&11.648&11.902&11.891&12.132&13.186&0.032\\
8&Reda Kriket&7.349&8.115&8.315&8.316&8.528&9.269&0.037\\
9&Ahmed Dahmani&7.390&7.982&8.133&8.129&8.274&8.922&0.027\\
10&Mohamed Abrini&6.242&6.810&6.943&6.949&7.084&7.730&0.029\\
			\hline 
		\end{tabular}}
		\caption{Statistical summary of the $1000$ estimations of the Banzhaf value for the $mwconn$ game.}
		\label{rankW}
	\end{center}
\end{table}
Table \ref{rankW} summarizes, from a purely statistical point of view, the $1000$ obtained results for the 10 more relevant terrorists in Zerkani network by using the estimated Banzhaf value with $\ell=10^3$ of the monotonic TU-game. Notice that the order established for the top-10 in  Table \ref{rank7_Banzhaf} can be also maintained  when using as criteria the main Statistical measures.

Analogous conclusions can be naturally obtained from the case of the additive TU-game in view of the Statistical summary for the $1000$ estimations of the Banzhaf value for the $10$ terrorists in the top-10 in the ranking of Table \ref{rank7_Banzhaf}. The numerical results are included in Table \ref{rankA}.

\begin{table}[h!]
	\begin{center}\resizebox{0.9\textwidth}{!}{
	\begin{tabular}{ p{0.3cm}| p{4cm}| p{1cm} p{1.1cm} p{1cm}  p{0.9cm} p{1.15cm} p{1 cm}| p{0.6cm} }
			\hline
			& Terrorist&Min.&1st Qu.&Median&Mean&3rd Qu.&Max. & CV  \\			\hline \hline
1&Mohamed Belkaid&29.127&30.907&31.386&31.412&31.896&33.714&0.023\\
2&Salah Abdeslam&23.860&25.678&26.182&26.167&26.679&28.434&0.028\\
3&Khalid Zerkani&24.599&25.440&25.727&25.716&25.956&26.857&0.014\\
4&Abdelhamid Abaaoud&23.080&23.952&24.157&24.156&24.363&25.172&0.013\\
5&Mohamed Bakkali&16.420&17.873&18.294&18.298&18.726&20.494&0.034\\
6&Najim Laachraoui&14.940&16.931&17.300&17.311&17.713&19.119&0.037\\
7&Fabien Clain&15.875&16.395&16.514&16.513&16.619&17.042&0.010\\
8&Reda Kriket&10.200&10.519&10.624&10.625&10.733&11.132&0.015\\
9&Mohamed Abrini&5.836&6.186&6.256&6.258&6.324&6.682&0.018\\
10&Miloud F.&5.492&5.765&5.833&5.833&5.894&6.136&0.017\\\hline
		\end{tabular}}
		\caption{Statistical summary of the $1000$ estimations of the Owen value for the $awconn$ game.}
		\label{rankA}
	\end{center}
\end{table}

From these results, we can check also the variability of rankings. From a purely quantitative approach, we compute the coefficient of variation (CV), that is,  the ratio of the standard deviation of the estimations and its mean. The proximity of their values to zero ensures the low numerical variability of the results, as well as the representativeness of the mean value. From a qualitative point of view,   Najim Laachraoui and Mohamed Bakkali, and Reda Kriket and Ahmed Dahmani, respectively reverse their positions under the monotonic perspective when considering the minimum estimations. Besides,  Fabien Clain and Najim Laachraoui also exchange their positions when considering the minimum values for the $awconn$ game.

We complete this analysis measuring the computational effort required in obtaining these $1000$ estimations. Table \ref{summary_timesB} includes a summary of the processing times (in seconds) that \verb|proc.time()| function in R software provides for each of these $1000$ repetitions. We distinguish between the User time, the System time and the Elapsed time. Focusing on the User time, we see that more than 75\% of the estimations have been obtained in less than 3 hours of real computing time.
\begin{table}[h!]
	\begin{center}\resizebox{0.8\textwidth}{!}{
		\begin{tabular}{ p{2cm} p{1.2cm} p{1.2cm} p{1.2cm}  p{1.2cm} p{1.2cm} p{1.5 cm} }
			\hline
			&Min.&1st Qu.&Median&Mean&3rd Qu.&Max. \\			\hline \hline
			User time &5385.142 & 8715.153 &9481.655& 9538.363& 10498.958 &13820.818
\\
			System time   &  4.745  &  6.393&    7.842&   23.041&    14.366&   330.296\\
			Elapsed time & 5390.902& 8721.884 &9490.329& 9562.124& 10547.094& 13848.808 \\
			\hline 
		\end{tabular}}
		\caption{Statistical summary of the $1000$ processing times (in seconds) for the estimations of the Banzhaf value.}
		\label{summary_timesB}
	\end{center}
\end{table}

\vspace*{-1 cm}
\subsection{The Banzhaf-Owen value approximation in Zerkani network}

Along this section, we analogously determine the $top-10$ of the ranking according to the relevance of the terrorists given by the estimation of the Banzhaf-Owen value in the Zerkani network. This section describes the R code specifically built for  applying the second procedure in Section \ref{sectionEBOwen} on the specific case of the Zerkani network.

\subsubsection{Definition of partition P in specialities}
From the data about members of the Zerkani network, the terrorists of the network have been grouped according to their rank, their function, their direct relationship with some terrorists, among others, in such a way that it gives rise to the creation of a partition with ten unions. That is, such coalition structure has ten unions, $P=\{P_{1},P_2,\cdots,P_{10}\}$ that are presented in Table \ref{partition}.

\begin{table}[h!]
	\begin{center}\resizebox{0.9\textwidth}{!}{
		\begin{tabular}{ p{0.8cm} p{12.2cm} }
			\hline
		Union	&Members \\			\hline \hline
		$P_{1}$ & Abdelhamid Abaaoud, Fabien Clain and Khalid Zerkani \\
	
	 $P_{2}$ & Chakib Akrouh, Gelel Attar, Hasna Ait Boulahcen, Fatima Aberkan, Osama Krayem, Souleymane Abrini, Ayoub el Khazzani, Mehdi Nemmouche, Thomas Mayet, Macreme Abrougui, Ahmed Dahmani and Adrien Guihal.\\
	
 $P_{3}$ & Sid Ahmed Ghlam, Reda Hame, AQI, Ilias Mohammadi, Soufiane Alilou, Najim Laachraoui and Khalid El Bakraoui.\\
	
$P_{4}$ & Paris Attacker A, Paris Attacker B, Salzburg Refugee A and Salzburg Refugee B.\\

$P_{5}$ & Mohammed Amri, Hamza Attou, Mohamed Abrini and Abid Aberkan. \\

$P_{6}$ & Mohamed Belkaid, Salah Abdeslam, Mohamed Bakkali and Ibrahim El Bakraoui.\\

$P_{7}$ & Reda Kriket, Rabah M., Y. A., Abderrahmane Ameroud, Miloud F., Anis Bari and AQIM.\\

$P_{8}$ & Khaled Ledjeradi and Djamal Eddine Ouali.\\

$P_{9}$ & Unknown identity and Tawfik A.\\
	
$P_{10}$& Ibrahim Abdeslam and Ali Oulkadi.\\
			\hline 
		\end{tabular}}
		\caption{Members of unions of the partition $P$ considered.}
		\label{partition}
	\end{center}
\end{table}

To introduce this information in R software we build a numerical vector  for the list of terrorists belonging to Zerkani network as follows.
\begin{example}
indexP<-rep(0,n)
indexP[c(1,9,15)]<-1                                           #Comanders and recruiters
indexP[c(8,11,12,10,22,28,32,33,29,35,5,4)]<-2     #Associates of commander or recruiter
indexP[c(31,34,39,13,41,21,44)]<-3        #Recruited or under the command of a commander
indexP[c(23,24,36,37)]<-4                             #Terrorists that traveled together
indexP[c(20,47,17,3)]<-5                              #They traveled with Salah Abdeslam
                                                   #or they had direct relation with him
indexP[c(19,27,18,43)]<-6                              #Traveled, lived together or were 
                                                     #associated before the Paris attack
indexP[c(26,25,30,2,16,7,42)]<-7                      #Members of the Reda Kriket's cell
indexP[c(14,40)]<-8                                          #Associated in Forging ring
indexP[c(46,45)]<-9                                                  #Arrested in Forest
indexP[c(38,6)]<-10                                                          #Associated
			\end{example}
For each component $\verb|i|$ of the vector \verb|indexP|, we specify the number of partition to which player $i$ belongs according to the imposed classification of the members of Zerkani network.

\subsubsection{R code}

Below, we depict the R code required to rank the terrorists belonging to Zerkani network based on the estimated Banzhaf-Owen value.

Once the auxiliary function is described, we focus on the study of \verb|estimate_BzOW()| function, that is, the R function built with the only purpose of estimating the Banzhaf-Owen value in the Zerkani network. This function has associated the following arguments as inputs:
\begin{itemize}
	\item \verb|seed|: a seed specifies the start point when a computer generates a random collection of permutations. That is, we can replicate the results when using the function in any computer and with the same value of the seed for a given set of parameters.
	\item \verb|nitmaxR|: a natural number which indicates the  number of coalitions of unions used in the estimation. If  \verb|nitmaxR| is larger than $2^{m-1}$, we internally reduce this amount to $2^{m-1}$.
	\item \verb|nitmaxS|: a natural number which indicates the  number of coalitions extracted from the union to each player belongs (that does not contain it) and that is considered for the estimation. If  \verb|nitmaxS| is larger than $2^{p_i-1}$, we internally reduce this amount to $2^{p_i-1}$.
	\item \verb|indexP|: a vector that indicates the union of partition $P$ to which each agent $i$ belongs.
	\item \verb|Individuals|: a vector that contains the individual weights of the terrorists involved in Zerkani network for measuring the influence of each of them.
	\item \verb|Zgraph|: a \verb|igraph| object that contains all the information associated to the graph induced by Zerkani network through the system of partitions that describe the possible affinities among its members.
	\item \verb|ZedgesNumber|: a \verb|data.frame| object that describes the edges of the graph in terms of nodes as well as its associated weight.
\end{itemize}

As output, this function also returns a list containing the following elements:
\begin{itemize}
	\item An estimation of the Banzhaf-Owen value for the TU-game $(N,v^{\mbox{awconn}})$ according to $P$. 
	\item An estimation of the Banzhaf-Owen value for the TU-game $(N,v^{\mbox{mwconn}})$ according to $P$. 
	\item Information about the processing times (in seconds) in terms of the User time, System time and Elapsed time that R software provides by using the basic function of R \verb|proc.time()|.
\end{itemize}

An example of its implementation in R software is displayed below:

\begin{example}
estimate_BzOW<-function(seed,nitmaxR,nitmaxS,indexP,Individuals,Zgraph,ZedgesNumber){
	
	set.seed(seed)
	n<-length(indexP); dim<-n
	nitmaxR0<-nitmaxR; nitmaxS0<-nitmaxS
	
	BzOA<-rep(0,n); BzOW<-rep(0,n)
	partition<-list()
        for (k in 1:max(indexP)){partition[[k]]<-which(indexP==k)}
    
	start<-proc.time()
	for (i in 1:n){
		
		ipartition<-indexP[i]
		mpartition<-partition[[ipartition]]
		iagent<-which(mpartition==i)
		pk<-length(mpartition)
		m<-max(indexP)
		
		nitmaxR<-min(nitmaxR0,2^(m-1))
		nitmaxS<-min(nitmaxS0,2^(pk-1))
		
		contP<-0
				
		ContributionW<-0
		ContributionA<-0
		while(contP<nitmaxR){
			Xm<-c()
			if (m>1){
				Xm<-sample(c(0,1),m,replace=T)
				Xm[ipartition]=0
			} else {
				Xm[1]<-c(1)
			}
			R<-c()
			im<-which(Xm>0);iml<-length(im)
			if (iml>0){
				for (k in 1:iml){R<-c(R,partition[[im[k]]])}
			}
			contP=contP+1
			contS<-0
			
			ContributionWS<-0
			ContributionAS<-0
			while(contS<nitmaxS){
				RS<-R
				Xpk<-sample(c(0,1),pk,replace=T)
				Xpk[iagent]<-0
				ipk<-which(Xpk>0)
				ipkl<-length(ipk)
				if (ipkl>0){
					for (k in 1:iml){RS<-c(RS,mpartition[which(Xpk>0)])}
				}
				RSui<-c(RS,i)
				
				if (length(RSui)==1){
					valueindividual=Individuals[RSui]
					ContributionWS=ContributionWS+valueindividual
					ContributionAS=ContributionAS+valueindividual
				} else {
					Zgraphwithi=induced_subgraph(Zgraph,RSui)
					Zgraphwithouti=induced_subgraph(Zgraph,RS)
					valuewith=mawconn(Zgraphwithi,Individuals,ZedgesNumber)
					valuewithout=mawconn(Zgraphwithouti,Individuals,ZedgesNumber)
					ContributionWS=ContributionWS+valuewith[1]-valuewithout[1]
					ContributionAS=ContributionAS+valuewith[2]-valuewithout[2]
				}
				contS=contS+1
			} 
			ContributionW=ContributionW+ContributionWS/contS
			ContributionA=ContributionA+ContributionAS/contS
		}
		BzOW[i]<-ContributionW/contP
		BzOA[i]<-ContributionA/contP
		
	}
	end<-proc.time()-start
	return(list(BzOA,BzOW,as.vector(end)[1:3]))
}
\end{example}

The lines of R code displayed below are considered as example of usage. It provides  an estimation of the Banzhaf-Owen value with sampling sizes equal to $\ell_r=10^2$ and $\ell_s=10$, respectively.
\begin{example}
BOW_Z<-estimate_BzOW(1,10^2,10,indexP,Individuals,Zgraph,ZedgesNumber)
\end{example}
Again, we can state that the existence of various processors  ensures a considerable diminishing of the computational complexity and the required computation time for its obtaining.

To complete this description, we show the usage of the \verb|ranking()| function in R software for ranking  terrorists  in Zerkani network according to the estimated Banzhaf-Owen value.
\begin{example}
ranking(BOW_Z[[1]],completenames)
ranking(BOW_Z[[2]],completenames)
\end{example}

\subsubsection{Numerical results}

In accordance with the scheme followed for the estimation of the Banzhaf value, we rank the members of the Zerkani network according to Banzhaf-Owen value. We use the R code described in the previous section and, once this estimation is obtained, we order the terrorists according to its non-increasing order.

By simplicity, we obtain $1000$ estimations of the Banzhaf-Owen value for the TU-games $(N,v^{mwconn})$ and $(N,v^{awconn})$ by using the two-stage sampling procedure described in Section \ref{sectionEOwen} with $\ell_r=\min\{10^2,2^{m-1}\}$ and $\ell_s=\min\{10,2^{p_i-1}\}$.
Note that sample sizes are taken to ensure that they are always smaller than the corresponding population sizes, i.e. $2^{m-1}2^{p_i-1}$, for each $i\in N$. Despite we are dealing with finite populations, we approximate the Banzhaf-Owen value as the average of these $1000$ approximations. Table \ref{rank7_BOwen} enumerates the Top-10 of the most relevant members of  the Zerkani network, when ordering according to the estimated Banzhaf-Owen value. A detailed list containing all ranked members is depicted in the Appendix B, more concretely, in Table \ref{rank7_BOwen_complete}. Khalid Zerkani is also the most influential terrorist under the monotonic approach when using the Banzhaf-Owen value, and being the second one under the additive approach. Again, Abdelhamid Abaaoud occupies the second position under the monotonic approach and he goes to  the fourth one under additivity. Now, Salah Abdeslam is in the third position under the both approaches considered for the TU-games. Similar conclusions can be obtained for the remainder of terrorists.

\begin{table}[h!]
	\begin{center}\resizebox{0.9\textwidth}{!}{
		\begin{tabular}{ p{0.5cm}|p{4cm} p{2cm}|p{4cm} p{2cm}}
			\hline
			& \multicolumn{2}{c|}{Ranking $R_{mwconn}$} & \multicolumn{2}{c}{Ranking $R_{awconn}$}\\\hline
			Pos. & Terrorist & ${\overline{BzO}}$ & Terrorist & ${\overline{BzO}}$ \\
			\hline \hline
1&Khalid Zerkani&39.498143&Mohamed Belkaid&32.192285\\
2&Abdelhamid Abaaoud&35.775328& Khalid Zerkani&27.932848\\
3&Salah Abdeslam&33.400368& Salah Abdeslam&26.959565\\
4& Mohamed Belkaid&33.380580& Abdelhamid Abaaoud&25.322645\\
5& Mohamed Bakkali&22.755968&Mohamed Bakkali&22.471120\\
6&Fabien Clain&12.285880& Fabien Clain&15.886888\\
7& Ahmed Dahmani& 9.879967&Reda Kriket&10.833910\\
8& Reda Kriket& 9.176184&Miloud F.& 5.906707\\
9&Najim Laachraoui& 4.701998&Ahmed Dahmani& 5.735320\\
10& Mohamed Abrini& 4.589148& Khaled Ledjeradi& 5.374710\\
			\hline 
		\end{tabular}}
		\caption{Top-10 of the ranking of terrorists in the Zerkani network, according to the average of $1000$ estimations of the Banzhaf-Owen value for  $(N,v^{mwconn})$ and $(N,v^{awconn})$ with $\ell_r=10^2$ and $\ell_s=10$.}
		\label{rank7_BOwen}
	\end{center}
\end{table}

After the Banzhaf-Owen value estimation, we check the variability of our sampling proposal this practical situation through a small simulation study. By construction, we rank all the members of Zerkani network in Table \ref{rank7_BOwen} but, by simplicity, we only study those ones in the top of the list. For this purpose, we summarize  the $1000$ obtained results for their estimated Banzhaf-Owen value. Table \ref{rankWB} summarizes the $1000$ obtained results for the estimated Banzhaf-Owen value for the case of the monotonic game.

\begin{table}[h!]
	\begin{center}\resizebox{0.9\textwidth}{!}{
	\begin{tabular}{ p{0.3cm}| p{4cm}| p{1cm} p{1.1cm} p{1cm}  p{0.9cm} p{1.15cm} p{1 cm}| p{0.6cm} }
			\hline
			& Terrorist&Min.&1st Qu.&Median&Mean&3rd Qu.&Max. & CV  \\			\hline \hline	1&Khalid Zerkani&33.050&38.069&39.425&39.498&40.956&47.903 &0.056\\
	2&Abdelhamid Abaaoud&30.618&34.709&35.761&35.775&36.833&41.350&0.045\\ 
	3&Salah Abdeslam&27.026&32.232&33.344&33.400&34.560&39.090&0.054\\ 
	4&Mohamed Belkaid&27.204&32.280&33.386&33.381&34.467&38.894&0.049\\
	5&Mohamed Bakkali&14.74&21.052&22.688&22.756&24.503&32.908&0.117\\ 
	6&Fabien Clain&9.54&11.753&12.278&12.286&12.803&15.095&0.064\\
	7&Ahmed Dahmani&6.385&9.192&9.861&9.880&10.518&12.970&0.100\\ 
	8&Reda Kriket&5.792&8.658&9.161&9.176&9.734&11.516 & 0.092\\
	9&Najim Laachraoui&3.659&4.478&4.708&4.702&4.925&6.117& 0.071\\ 
	10&Mohamed Abrini&3.660&4.370&4.580&4.589&4.814&5.610&0.073\\ 
			\hline 
		\end{tabular}}
		\caption{Statistical summary of the $1000$ estimations of the Banzhaf-Owen value for the $mwconn$ game.}
		\label{rankWB}
	\end{center}
\end{table}

Table \ref{rankAB} summarizes, from a Statistical point of view, the $1000$ estimations of the Banzhaf-Owen value for the individuals in the top-10 when considering the additive game.

\begin{table}[h!]
	\begin{center}\resizebox{0.9\textwidth}{!}{
	\begin{tabular}{ p{0.3cm}| p{4cm}| p{1cm} p{1.1cm} p{1cm}  p{0.9cm} p{1.15cm} p{1 cm}| p{0.6cm} }
			\hline
			& Terrorist&Min.&1st Qu.&Median&Mean&3rd Qu.&Max. & CV  \\			\hline \hline
1&Mohamed Belkaid   &28.020  &31.270    &32.131&32.192 &33.158&37.035 & 0.043 \\
2&Khalid Zerkani    &23.765 &27.077&27.896&27.933&28.790    &32.045 &0.045 \\
3&Salah Abdeslam    &22.468&25.942&26.945  &26.960 &27.949 &31.808&0.053\\
4&Abdelhamid Abaaoud&22.575 &24.647&25.353 &25.323 &25.976&28.408 &0.039\\
5&Mohamed Bakkali   &15.015 &20.814&22.424&22.471  &24.100   &31.750 & 0.112  \\
6&Fabien Clain      &14.413&15.548  &15.889&15.887&16.183&17.803 & 0.031\\
7&Reda Kriket       &9.648  &10.579  &10.835 &10.834  &11.077 &12.056 & 0.035 \\
8&Miloud F.         &5.501  &5.813    &5.906   &5.907  &5.998   &6.297   & 0.023\\
9&Ahmed Dahmani     &4.200    &5.435    &5.720    &5.735   &6.040     &7.380  & 0.079 \\
10&Khaled Ledjeradi  &4.640   &5.240     &5.370    &5.375   &5.510     &6.050 & 0.040\\
			\hline 
		\end{tabular}}
		\caption{Statistical summary of the $1000$ estimations of the Banzhaf-Owen value for the $awconn$ game.}
		\label{rankAB}
	\end{center}
\end{table}

In view of the results, the rankings obtained generally hold even over the maximum and minimum values of the estimated components of the Banzhaf-Owen value  in both scenarios. The exception is the exchange of Abdelhamid Abaaoud and Mohamed Bakkali's positions when considering  the maximum estimations of the Banzhaf-Owen value for the $awconn$ game. Regarding to the coefficient of variation (CV), we observe smaller values, in general, under the additive approach.

Again, we analyse the processing times (in seconds) required in the Banzhaf-Owen value estimation as a measure of the effort for this purpose. They are described in Table \ref{summary_timesBOW} includes a summary of these amounts. 

\begin{table}[h!]
	\begin{center}\resizebox{0.8\textwidth}{!}{
		\begin{tabular}{ p{2cm} p{1.2cm} p{1.2cm} p{1.2cm}  p{1.2cm} p{1.2cm} p{1.5 cm} }
			\hline
			&Min.&1st Qu.&Median&Mean&3rd Qu.&Max. \\			\hline \hline
			User time & 4610.515& 7658.268& 8223.145 &8359.013 & 9203.457& 11959.200\\
				System time   &    2.790 &   4.208 &   5.298&   17.124&    8.905 &   260.98\\
			Elapsed time &4614.003 &7662.244& 8235.254& 8376.096& 9217.770& 12170.890\\
			\hline 
		\end{tabular}}
		\caption{Statistical summary of the $1000$ processing times (in seconds) for the estimations of the Banzhaf-Owen value.}
		\label{summary_timesBOW}
	\end{center}
\end{table}
\vspace*{-1 cm}

\section{Comparison with the usage of the Shapley value and the Owen value }\label{discussion}

In this section, we do a brief discussion on the rankings obtained under the estimations of the Banzhaf value and the Banzhaf-Owen value in comparison to the ones under the Shapley value and the Owen value, respectively. To this purpose, we use the sampling procedures considered in \cite{Castro:2009} and \cite{Saavedra:2018} that, based on simple random sampling with replacement, respectively provide approximations of the two mentioned values.  By simplicity, and for the sole purpose of comparing the scenarios,  we estimate $100$ times the Shapley value and the Owen value by taking $\ell=1000$ permutations of $N$. These estimations will be denoted by ${\overline{Sh}}$ and ${\overline{O}}$, respectively. The complete list of terrorists based on these results as well as the R-code required for their obtaining can be supplied on request. For the case of the Banzhaf value, we  consider the average of  $100$ estimations by taking $\ell=1000$ coalitions of $N\setminus \{i\}$, for each $i\in N$, and the average of $100$ estimations of the Banzhaf-Owen value with $\ell_r=10^2$ and $\ell_s=10$. Table \ref{rank:comparison} illustrates the top-10 of the rankings obtained. 

\begin{table}[h!]
	\begin{center}\resizebox{1.05\textwidth}{!}{
		\begin{tabular}{|p{2.795 cm} p{0.8 cm}|p{2.795 cm} p{0.8 cm}||p{2.795 cm} p{0.8 cm}|p{2.795 cm} p{0.8 cm}|}
			\hline
			 \multicolumn{4}{|c||}{Ranking $R_{mwconn}$} & \multicolumn{4}{c|}{Ranking $R_{awconn}$}\\\hline \hline
			 Terrorist & ${\overline{Sh}}$ & Terrorist & ${\overline{O}}$ & Terrorist & ${\overline{Sh}}$ & Terrorist & ${\overline{O}}$ \\
			\hline \hline
Ab. Abaaoud& 17.108& Khalid Zerkani& 39.242&Mohamed Belkaid& 13.987&Mohamed Belkaid& 28.460\\
Khalid Zerkani& 15.026 & Ab. Abaaoud& 36.129& Khalid Zerkani& 12.332& Khalid Zerkani& 27.677\\
Salah Abdeslam& 14.741&Mohamed Belkaid& 29.236 &Ab. Abaaoud& 11.850&Mohamed Bakkali& 27.168\\
Mohamed Belkaid& 14.249&Mohamed Bakkali& 27.845& Salah Abdeslam& 11.453& Ab. Abaaoud& 26.157\\
Najim Laachraoui&7.918& Salah Abdeslam& 27.042& Fabien Clain&8.295& Salah Abdeslam& 22.439\\
Mohamed Bakkali&7.356& Fabien Clain& 13.661&Mohamed Bakkali&7.625& Fabien Clain& 16.404\\
Fabien Clain&5.884&Reda Kriket& 11.642& Najim Laachraoui&7.549&Reda Kriket& 11.395\\
Reda Kriket&3.696&Ahmed Dahmani& 10.754&Reda Kriket&4.923&Ahmed Dahmani&6.142\\
Ahmed Dahmani&3.369& Khaled Ledjeradi&4.702& Mohamed Abrini&2.996&Miloud F.&6.093\\
Mohamed Abrini&2.917&Miloud F.&4.209&Miloud F.&2.827& Khaled Ledjeradi&5.271\\
\hline\hline
	 Terrorist & ${\overline{Bz}}$ & Terrorist & ${\overline{BzO}}$ & Terrorist & ${\overline{Bz}}$ & Terrorist & ${\overline{BzO}}$ \\\hline\hline
Ab. Abaaoud&38.372&Khalid Zerkani&39.328&Mohamed Belkaid&31.333&Mohamed Belkaid&32.274\\
Salah Abdeslam&34.993&Ab. Abaaoud&35.702&Salah Abdeslam&26.114&Khalid Zerkani&27.854\\
Khalid Zerkani&33.992&Salah Abdeslam&33.639&Khalid Zerkani&25.752 & Salah Abdeslam&27.010\\
Mohamed Belkaid&33.144&Mohamed Belkaid&33.400&Ab. Abaaoud &24.206&Ab. Abaaoud&25.426\\
Najim Laachraoui&18.827&Mohamed Bakkali&22.473&Mohamed Bakkali&18.360&Mohamed Bakkali&22.181\\
Mohamed Bakkali&18.367&Fabien Clain&12.298&Najim Laachraoui&17.381&Fabien Clain&15.892\\
Fabien Clain&11.903&Ahmed Dahmani&9.900&Fabien Clain&16.538&Reda Kriket&10.830\\
Reda Kriket&8.316&Reda Kriket&9.166&Reda Kriket&10.620&Miloud F. &5.916\\
Ahmed Dahmani&8.111&Najim Laachraoui&4.740&Mohamed Abrini&6.242&Ahmed Dahmani&5.712\\
Mohamed Abrini&6.924&Mohamed Abrini&4.621&Miloud F. &5.833&Khaled Ledjeradi&5.379\\
			\hline 
		\end{tabular}}
		\caption{Top-10 of terrorists in Zerkani network, according to the average of $100$ estimations of the Owen value and the Shapley value with $\ell=1000$ (top), of the average of $100$ estimations of the Banzhaf value with $\ell=1000$, and of the average of $100$ estimations of the Banzhaf-Owen value with $\ell_r=10^2$ and $\ell_s=10$ and of the Banzhaf value with $\ell=1000$ (bottom).}
		\label{rank:comparison}
	\end{center}
\end{table}

Below, we briefly do some comments on the resulting rankings of the members of Zerkani network. In general, no changes are remarkable in the list of individuals belonging to the top-10 in the obtained rankings although we can emphasize the fact of the most of the positions change when using the different approaches considered in this paper. Probably, this may be due to the organisational and logistical role played by such terrorists in such a way that their weight will change with the approach under consideration. Notice that, others, as the ones who carry out the suicide and therefore the action (Salzburg Refugee A or B, among others) usually appear from eleventh position onwards.

Khalid Zerkani is leading the rankings when a priori unions system exist under the monotonic approach, with respect to the ones associated to the cases of the Shapley and the Banzhaf values. Recall that this man who directed a recruitment network in the Brussels area. He was not present or coordinated the attacks of Paris and Brussels, but had a high influence on all those who related to him. He is currently imprisoned on terrorism related charges. In general, we check that Khalid Zerkani generally moves up positions in the top-10 under the presence of a priori union systems except in the case of the Shapley value for the additive approach, where he maintains the same position. In this additive scenario, the person who always occupies the first position in all rankings is Mohamed Belkaid, even without a priori unions systems. Najim Laachraoui and Mohamed Abrini did not belong to the top-10 under the usage of the Owen value in ranking, when monotonic and additive approaches are assumed, as well as the Banzhaf-Owen value is considered under the additive TU-game. Their positions are completed with Khaled Ledjeradi and Miloud F., when the Owen value is considered for the monotonic TU-game. However, under the additive approach,  Najim Laachraoui and Mohamed Abrini are changed by Ahmed Dahmani and Khaled Ledjeradi in the lists.

In general, these results are in line with the reality of the Zerkani network. Along with Khalid Zerkani and Abdelhamid Abaaoud, the role of Mohamed Bakkali is also important since that it alleged intellectual author of the attack of Paris. It is believed that he selected those who were going to be in the war zones or in Europe. He died in a police raid, when Chakib Akrouh detonated his explosives belt. Another individual in the ranking is Salah Abdeslam. He was the most wanted man in Europe after the Paris attack. Fabien Clain was one planner of Paris attack and explored the different places where to perform the blows. Then, Reda Kriket was a recruiter for the network and provided money to it. Meanwhile, Khaled Ledjeradi was someone very required in the network, since he was the leader of an organization that created fake documents for the members of the network, allowing them to travel. Finally, about Miloud F. not much information is available, but he was arrested in Turkey in 2005 and this allowed for arresting Reda Kriket later. About Salzburg Refugee A and Salzburg Refugee B refugees, barely there is information about them, but they are the points of union of the attackers Paris A and B to the network, so the supervision of the last two subjects mentioned may have been key to the cessation of the attack. The decision not to increase police surveillance on them, even if it was not a good one, can be justified from our results since only under the additive perspective, these individuals rank high in the ranking.

Finally, we compare the overall rankings by using graphical tools. To this purpose, the Lorenz curve is a graphical representation frequently used to show the relative distribution of a variable (the solution concept) in a given domain (given in this case by the set of terrorists). The curve is plotted considering on the horizontal axis the cumulative percentage of people and on the vertical axis the cumulative percentage of the variable.  In terms of comparing two Lorenz curves, if one  is always above the other (hence closer to the 45 degree line than the other), then we can unambiguously say that the former exhibits less inequality than the latter.

\begin{figure}[h!]
	{\centering
		\includegraphics[height=7 cm,width=0.3\textheight]{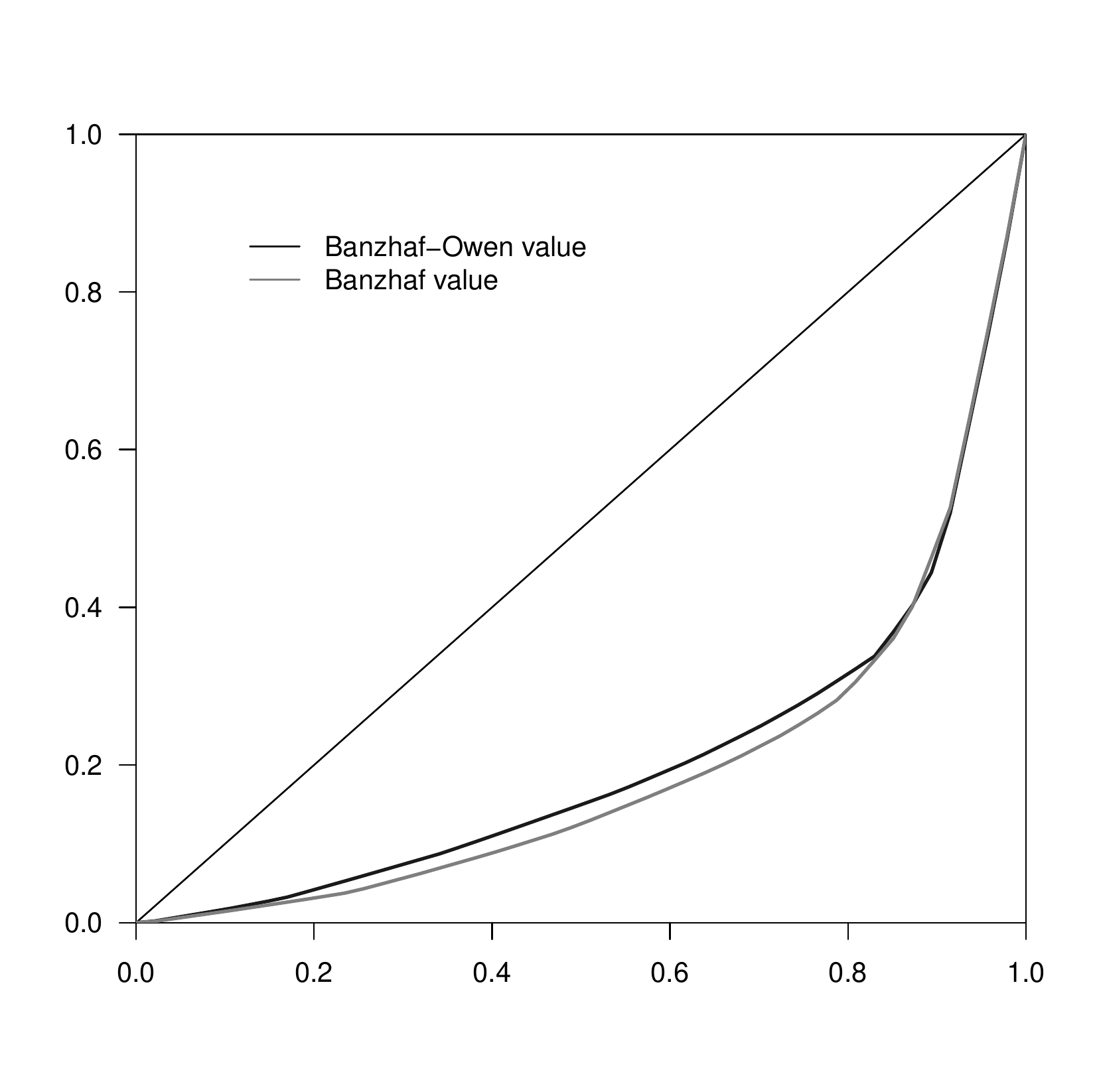}
		\includegraphics[height=7 cm,width=0.3\textheight]{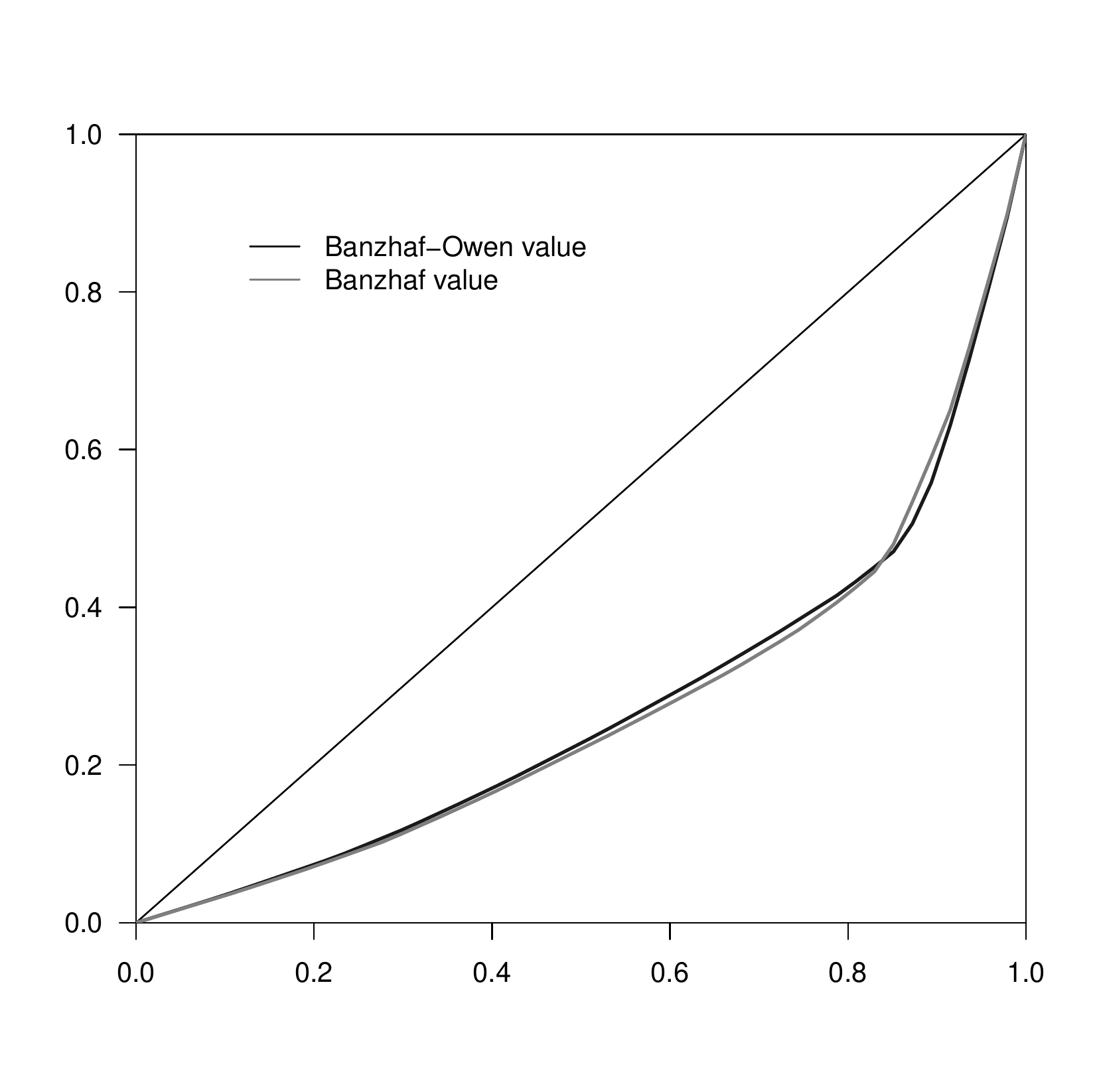}
	}
	{\centering
		\includegraphics[height=7 cm,width=0.3\textheight]{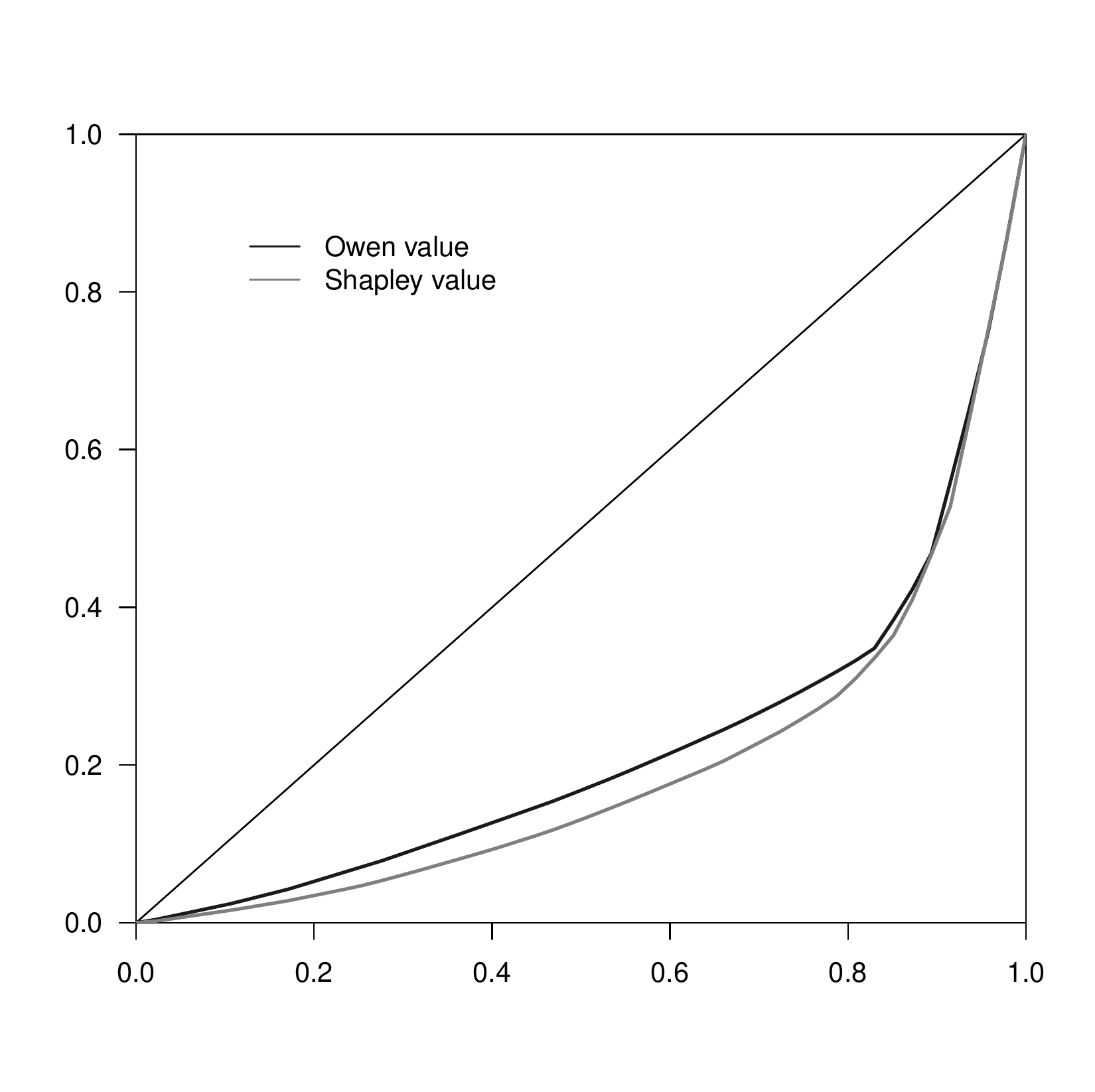}
		\includegraphics[height=7 cm,width=0.3\textheight]{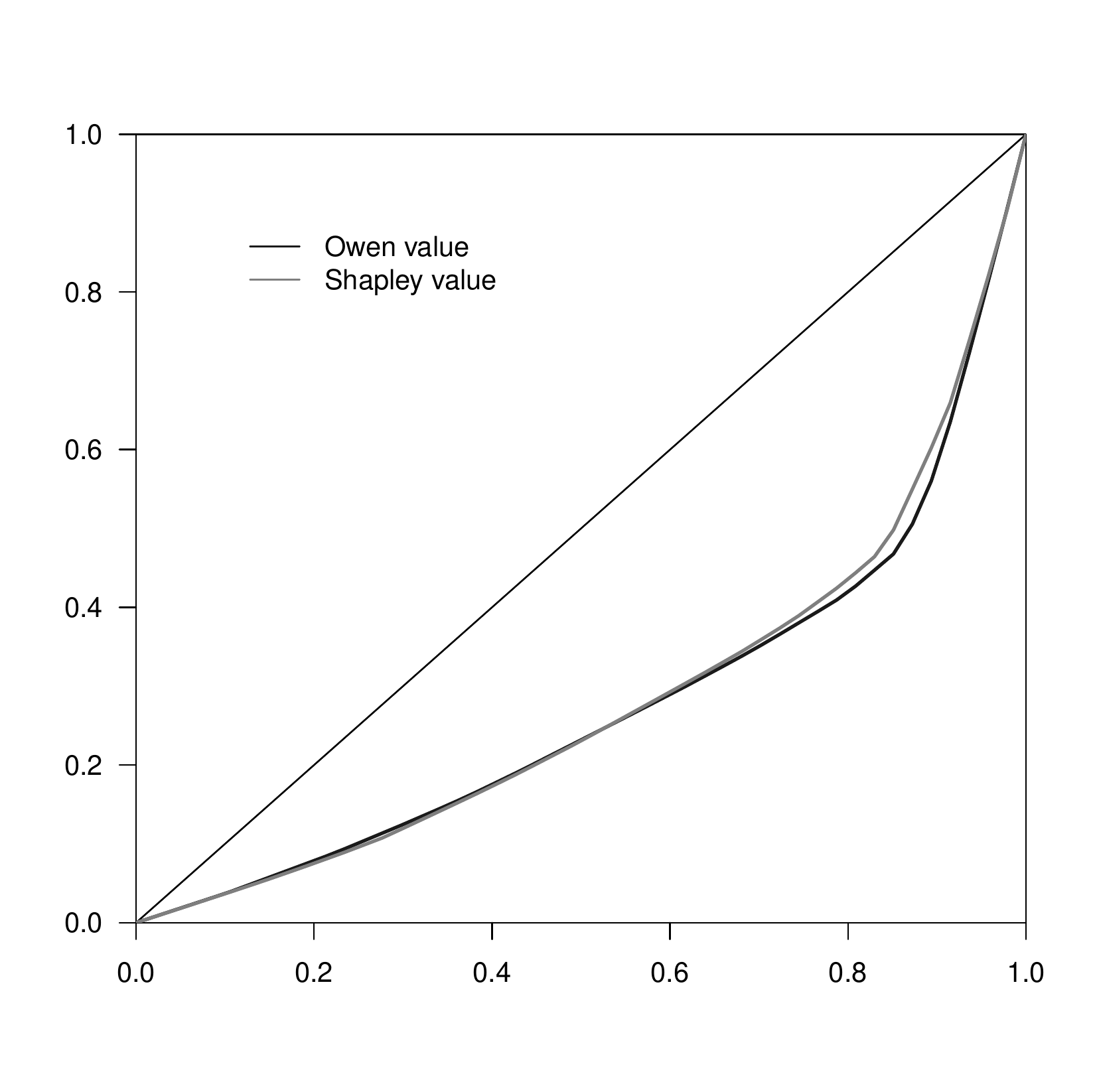}
   }
	
	\begin{picture}(1,10)
	
	\put(-10,240){\begin{sideways}Banzhaf-Owen value and Banzhaf value\end{sideways} }
	\put(-10,55){\begin{sideways}Owen value and Shapley value\end{sideways} }	
	\put(90,410){$(N,v^{mwconn})$}
	\put(300,410){$(N,v^{awconn})$}
	\end{picture}
	
	\vspace*{-1.3 cm}\caption{Lorenz curves comparison of $100$ estimations of the Owen value and the Shapley value with $\ell=1000$ (top) for $(N,v^{mwconn})$ (left) and $(N,v^{awconn})$ (right), and for $100$ estimations of the Banzhaf-Owen value with $\ell_r=10^2$ and $\ell_s=10$ and of the Banzhaf value with $\ell=1000$ (bottom) for $(N,v^{mwconn})$ (left) and $(N,v^{awconn})$ (right).}
	\label{fig:lorenz_curves} 
\end{figure}
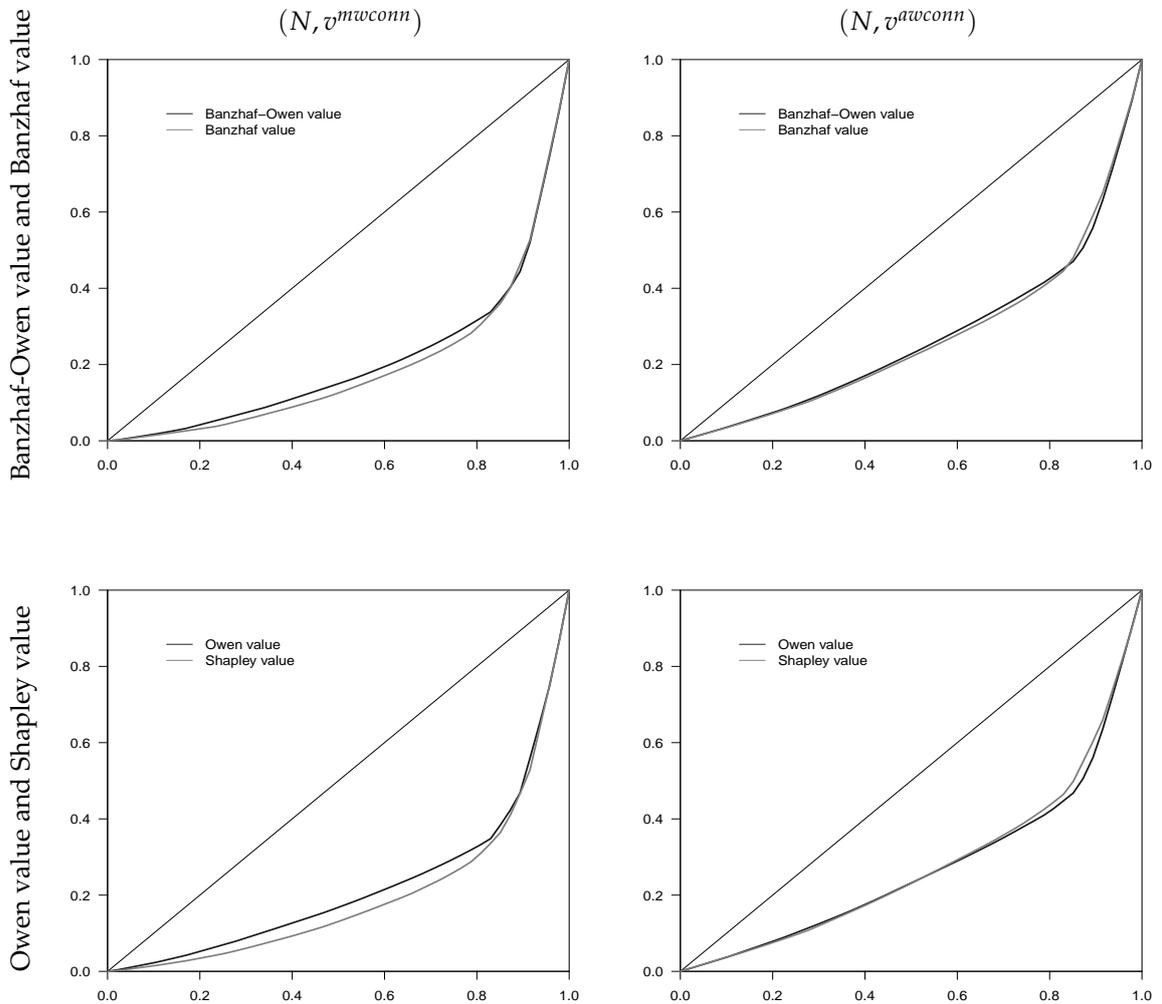

	We plot the empirical Lorenz curve of the estimated Banzhaf values and of the estimated Banzhaf-Owen values for the monotonic and the additive TU-games considered, respectively. We also compare with the associated curves to the estimated Shapley value and the estimated Owen value. To this aim, we use \code{Lc()} function in \pkg{ineq} package in R software. Figure \ref{fig:lorenz_curves} shows the associated results. 
	
	From the graphical representations, we can ensure that, for $(N,v^{mwconn})$, the Owen value and the Banzhaf-Owen value estimations show smaller inequalities between components than the corresponding Shapley and Banzhaf value estimations. However, under the additive approach, there are no apparent differences between the two estimations since that the two curves coincide in practice.

\section{Conclusions}

Until now, the usage of game-theoretical solutions for the analysis of terrorist networks was limited to the approach given by the Shapley value. However, there exist other ones, based on different criteria, that have not been considered for this purpose.  In this paper, the Banzhaf value and its extension for the case in which a coalitional structure that restricts the affinities among their members exists, the Banzhaf-Owen value, have been innovatively considered to obtain rankings of the most important terrorist of the Zerkani network. By using these, new and interesting tools or measures of centrality are offered in these networks. Additionally, and according to the results obtained, it can be concluded that it can be useful and interesting for the intelligence services.
					
As in any organization in which many people are involved, the members of  a terrorist network are hierarchical within. Although all are part of the same network, each of them will perform a function, and it will be related to nearby members of those of the same rank. It makes sense, therefore, to consider the members of the network in groups according to their features.  This directly implies the existence of a partition, as the one considered along this paper. Thus, the construction of two rankings has been carried out through the approximation of the Banzhaf-Owen value, for the two approaches of TU-games under study. Both take into account the characteristics of the network, of the individuals and their relationships in a monotonic or additive form. One of the objectives was to compare the results obtained by the estimated solutions with those that \citet{Hamers:2019} get for the  Shapley value approximation in this setting as well as the approximation of the Owen value, its extension for the case with a coalition structure. In order to choose the proper partition to apply the Owen value and the Banzhaf-Owen value, a deeper analysis of each terrorist's contributions to the network has been made, having divided them according to their facets thanks to the consideration of a coalitional structure in the Owen and Banzhaf-Owen values. In this sense, the characteristics of the most realistic partition may be indicated by those responsible of the security services, since they usually manage relevant information on the organization of this kind of terrorist structures. Even so, the usage of more technical mechanisms to determine that partition to be considered may be addressed from a quantitative point of view.

Focus on the partition considered, notice that the commanders and recruiters within the network will be those who manage, order and coordinate all the movements that the group carries out. However, they will avoid getting involved in compromised issues due to their status. For this situation, they have associates or recruits who will obey them and serve as a connection with the attackers (which would be in a lowest level in the network). Meanwhile, lower-level groupings will carry out the hard work of exposing themselves to intelligence agencies, traveling throughout Europe with false documentation or carrying out the killing actions of hundreds of people. Unlike of the Shapley or now Banzhaf values, all this information about the partition is taken into account in the Owen value and the Banzhaf-Owen value. Therefore, the results integrates this additional information, which can be key in determining who are the most dangerous or influential members within each union or group considered in the partition. Specifically, the Owen and Banzhaf-Owen values introduce a coalitional structure that, respectively, enriches the results provided by the Shapley and Banzhaf values when extra information about relationships is considered. Both distribute first among unions (teams) and then among the members of each team. Therefore, when considering the terrorists according to their facets, the most dangerous in each team may be obtained and, in particular, not only the most influential of the organization or recruitment as the Shapley value and the Banzhaf value do, but also the most important among those who carry out the action, which may be essential to neutralize it.

\section*{Acknowledgements}
The authors wish to thank Herbert Hamers for his insightful comments on an earlier version of this work.\\
Financial support from R\&D\&I Project Grant PGC2018-097965-B-I00, funded by MCIN/ AEI/10.13039/\newline 501100011033/ and by ``ERDF A way of making Europe''/EU is gratefully acknowledged. A. Saavedra-Nieves acknowledges the financial support of FEDER/Ministerio de Ciencia, Innovaci\'on y Universidades - Agencia Estatal de Investigaci\'on under grant MTM2017-87197-C3-3-P, and of the Xunta de Galicia through the European Regional Development Fund (Grupos de Referencia Competitiva ED431C 2021/24). Authors also thank the computational resources of the Centro de Supercomputación de Galicia (CESGA).
	$ $
					
					\address{Encarnación Algaba\\
						Matemática Aplicada II and Instituto de Matemáticas de la Universidad de Sevilla (IMUS), Escuela Superior de Ingenieros, Camino de los Descubrimientos, s/n, 41092 Sevilla,\\
						Spain\\
						\email{ealgaba@us.es}}

					\address{Andrea Prieto\\
						Matemática Aplicada II, Escuela Superior de Ingenieros, Camino de los Descubrimientos, s/n, 41092 Sevilla,\\
						Spain\\
						\email{aprietogarcia@us.es}}
					
				\address{Alejandro Saavedra-Nieves\\
					Departamento de Estatística, Análise Matemática e Optimización, Facultade de Matemáticas, Universidade de Santiago de Compostela,\\
					Spain\\
					\email{alejandro.saavedra.nieves@usc.es}}
					
					\bibliography{algaba-prieto-saavedra}
					
					\clearpage\section{Appendix A. A ranking based on the Banzhaf value}\label{appA}
					
					Table \ref{rank7_Bz_complete} depicts the estimation of the Banzhaf value for the TU-games with a priori unions $(N,v^{mwconn})$ and $(N,v^{awconn})$  associated to Zerkani network with $\ell=10^6$.
					
	\begin{table}[h!]
		\begin{center}\resizebox{0.55\textheight}{0.7\textwidth}{
			\begin{tabular}{ p{0.5cm}|p{4cm} p{2cm}|p{4cm} p{2cm}}
				\hline
				& \multicolumn{2}{c|}{Ranking $R_{mwconn}$} & \multicolumn{2}{c}{Ranking $R_{awconn}$}\\\hline
				Pos. & Terrorist & $\overline{Bz}$ & Terrorist & $\overline{Bz}$ \\
				\hline \hline
1	&	Abdelhamid Abaaoud	&	38.326053	&	Mohamed Belkaid	&	31.411917	\\
2	&	Salah Abdeslam	&	35.073561	&	Salah Abdeslam	&	26.167231	\\
3	&	Khalid Zerkani	&	33.930235	&	Khalid Zerkani	&	25.716073	\\
4	&	Mohamed Belkaid	&	33.267557	&	Abdelhamid Abaaoud	&	24.155854	\\
5	&	Najim Laachraoui	&	18.774721	&	Mohamed Bakkali	&	18.297970	\\\hline
6	&	Mohamed Bakkali	&	18.307468	&	Najim Laachraoui	&	17.311352	\\
7	&	Fabien Clain	&	11.891287	&	Fabien Clain	&	16.512582	\\
8	&	Reda Kriket	&	8.316217	&	Reda Kriket	&	10.625338	\\
9	&	Ahmed Dahmani	&	8.129191	&	Mohamed Abrini	&	6.257625	\\
10	&	Mohamed Abrini	&	6.94882	&	Miloud F.	&	5.832491	\\
11	&	Khaled Ledjeradi	&	4.860392	&	Ahmed Dahmani	&	5.371277	\\
12	&	Ilias Mohammadi	&	4.419613	&	Khaled Ledjeradi	&	5.253377	\\
13	&	Souleymane Abrini	&	4.192845	&	Ilias Mohammadi	&	4.602826	\\
14	&	Chakib Akrouh	&	3.682052	&	Salzburg Refugee B	&	4.419092	\\
15	&	Hasna Ait Boulahcen	&	3.669759	&	Salzburg Refugee A	&	4.418852	\\
16	&	Gelel Attar	&	3.38936	&	Ali Oulkadi	&	4.195854	\\
17	&	Ali Oulkadi	&	3.24243	&	Chakib Akrouh	&	3.949340	\\
18	&	Abid Aberkan	&	3.059297	&	Khalid El Bakraoui	&	3.938956	\\
19	&	Mohammed Amri	&	3.045479	&	Hasna Ait Boulahcen	&	3.917760	\\
20	&	Osama Krayem	&	2.976607	&	Osama Krayem	&	3.875802	\\
21	&	Fatima Aberkan	&	2.884989	&	Souleymane Abrini	&	3.828365	\\
22	&	Ibrahim El Bakraoui	&	2.876029	&	Mohammed Amri	&	3.826162	\\
23	&	Mehdi Nemmouche	&	2.844589	&	Gelel Attar	&	3.668426	\\
24	&	Khalid El Bakraoui	&	2.723389	&	Macreme Abrougui	&	3.660526	\\
25	&	Miloud F.	&	2.514147	&	Adrien Guihal	&	3.660303	\\
26	&	Salzburg Refugee B	&	2.235932	&	Thomas Mayet	&	3.658254	\\
27	&	Salzburg Refugee A	&	2.231646	&	Mehdi Nemmouche	&	3.620552	\\
28	&	Ibrahim Abdeslam	&	2.195443	&	Abid Aberkan	&	3.594534	\\
29	&	Hamza Attou	&	2.103029	&	Fatima Aberkan	&	3.450278	\\
30	&	Sid Ahmed Ghlam	&	2.026433	&	Ibrahim Abdeslam	&	3.382164	\\
31	&	Identity Unknown	&	2.026294	&	Ibrahim El Bakraoui	&	3.349120	\\
32	&	Tawfik A.	&	2.024749	&	Paris Attacker B	&	3.271808	\\
33	&	Soufiane Alilou	&	1.919801	&	Paris Attacker A	&	3.270114	\\
34	&	Ayoub el Khazzani	&	1.899183	&	Sid Ahmed Ghlam	&	3.120720	\\
35	&	Reda Hame	&	1.895839	&	Identity Unknown	&	2.656934	\\
36	&	Djamal Eddine Ouali	&	1.635981	&	Tawfik A.	&	2.655042	\\
37	&	Paris Attacker B	&	1.130785	&	Hamza Attou	&	2.624632	\\
38	&	Paris Attacker A	&	1.125123	&	Djamal Eddine Ouali	&	2.562870	\\
39	&	Macreme Abrougui	&	1.079897	&	Soufiane Alilou	&	2.445775	\\
40	&	Thomas Mayet	&	1.078904	&	Ayoub el Khazzani	&	2.414432	\\
41	&	Adrien Guihal	&	1.078749	&	Reda Hame	&	2.408251	\\
42	&	AQI	&	1.064913	&	AQIM	&	2.359843	\\
43	&	Anis Bari	&	1.009151	&	Anis Bari	&	2.252614	\\
44	&	Rabah M.	&	1.004866	&	Abderrahmane Ameroud	&	2.249956	\\
45	&	Abderrahmane Ameroud	&	1.004735	&	Rabah M.	&	2.249280	\\
46	&	Y. A.	&	1.002506	&	Y. A.	&	2.248912	\\
47	&	AQIM	&	0.515864	&	AQI	&	2.248712	\\

				\hline 
			\end{tabular}}
			\caption{Overall ranking of terrorists in the Zerkani network, according to the estimated Banzhaf value for games $(N,v^{mwconn})$ and $(N,v^{awconn})$ with $\ell=10^6$.}
			\label{rank7_Bz_complete}
		\end{center}
	\end{table}
					
					\clearpage
\section{Appendix B. A ranking based on the Banzhaf-Owen value}\label{appB}

Table \ref{rank7_BOwen_complete} depicts the estimation of the Banzhaf-Owen value for the TU-games with a priori unions $(N,v^{mwconn})$ and $(N,v^{awconn})$  associated to Zerkani network with $\ell_r=10^2$ and $\ell_s=10$.
					
\begin{table}[h!]
	\begin{center}\resizebox{0.55\textheight}{0.7\textwidth}{
		\begin{tabular}{ p{0.5cm}|p{4cm} p{2cm}|p{4cm} p{2cm}}
			\hline
			& \multicolumn{2}{c|}{Ranking $R_{mwconn}$} & \multicolumn{2}{c}{Ranking $R_{awconn}$}\\\hline
			Pos. & Terrorist & ${\overline{BzO}}$ & Terrorist & ${\overline{BzO}}$ \\
			\hline \hline
			1&Khalid Zerkani&39.498143&Mohamed Belkaid&32.192285\\
			2&Abdelhamid Abaaoud&35.775328& Khalid Zerkani&27.932848\\
			3&Salah Abdeslam&33.400368& Salah Abdeslam&26.959565\\
			4& Mohamed Belkaid&33.380580& Abdelhamid Abaaoud&25.322645\\
			5& Mohamed Bakkali&22.755968&Mohamed Bakkali&22.471120\\\hline
			6&Fabien Clain&12.285880& Fabien Clain&15.886888\\
			7& Ahmed Dahmani& 9.879967&Reda Kriket&10.833910\\
			8& Reda Kriket& 9.176184&Miloud F.& 5.906707\\
			9&Najim Laachraoui& 4.701998&Ahmed Dahmani& 5.735320\\
			10& Mohamed Abrini& 4.589148& Khaled Ledjeradi& 5.374710\\
			11& Khaled Ledjeradi& 4.543365& Salzburg Refugee A& 4.635760\\
			12&Hasna Ait Boulahcen& 4.291352& Salzburg Refugee B& 4.627397\\
			13&Ilias Mohammadi& 4.006257&Ali Oulkadi& 4.588490\\
			14&Souleymane Abrini& 3.876333&Ilias Mohammadi& 4.407872\\
			15&Ali Oulkadi& 3.573005&Hasna Ait Boulahcen& 4.388288\\
			16& Abid Aberkan& 3.521660& Mohamed Abrini& 4.306460\\
			17&Mohammed Amri& 3.506505&Mohammed Amri& 4.257300\\
			18&Chakib Akrouh& 3.277582&Chakib Akrouh& 4.078612\\
			19&Mehdi Nemmouche& 3.005955& Najim Laachraoui& 4.031124\\
			20& Fatima Aberkan& 3.004538& Abid Aberkan& 4.015080\\
			21&Miloud F.& 3.003710& Khalid El Bakraoui& 4.003000\\
			22&Ibrahim El Bakraoui& 2.798625& Osama Krayem& 3.998680\\
			23& Salzburg Refugee A& 2.583131&Souleymane Abrini& 3.893326\\
			24& Salzburg Refugee B& 2.561874&Mehdi Nemmouche& 3.755620\\
			25&Tawfik A.& 2.506115& Thomas Mayet& 3.755484\\
			26& Osama Krayem& 2.504972& Macreme Abrougui& 3.749120\\
			27& Khalid El Bakraoui& 2.502982&Adrien Guihal& 3.746474\\
			28&Hamza Attou& 2.498839& Ibrahim Abdeslam& 3.599290\\
			29& Identity Unknown& 2.491900& Fatima Aberkan& 3.500740\\
			30&Gelel Attar& 2.374725&Ibrahim El Bakraoui& 3.466547\\
			31& Ibrahim Abdeslam& 2.340145& Paris Attacker B& 3.376955\\
			32& Thomas Mayet& 2.036028& Paris Attacker A& 3.371785\\
			33& Macreme Abrougui& 2.028873&Gelel Attar& 3.242516\\
			34&Adrien Guihal& 2.016827&Tawfik A.& 3.004120\\
			35&AQI& 2.005373&Hamza Attou& 2.998880\\
			36&Reda Hame& 2.005209& Identity Unknown& 2.992720\\
			37&Sid Ahmed Ghlam& 1.998750&Djamal Eddine Ouali& 2.752420\\
			38&Ayoub el Khazzani& 1.994801&AQI& 2.504360\\
			39&Soufiane Alilou& 1.993469&Reda Hame& 2.504340\\
			40&Djamal Eddine Ouali& 1.529595&Sid Ahmed Ghlam& 2.498860\\
			41& Paris Attacker A& 1.340002&Soufiane Alilou& 2.494900\\
			42& Paris Attacker B& 1.337836&Ayoub el Khazzani& 2.494580\\
			43&Y. A.& 1.195130& AQIM& 2.376069\\
			44&Anis Bari& 1.190284&Y. A.& 2.283428\\
			45& Rabah M.& 1.188459&Anis Bari& 2.281672\\
			46& Abderrahmane Ameroud& 1.184284 & Abderrahmane Ameroud& 2.280678\\
			47& AQIM& 0.654073& Rabah M.& 2.279296\\
			\hline 
		\end{tabular}}
		\caption{Overall ranking of terrorists in the Zerkani network, according to the estimated Banzhaf-Owen value for games $(N,v^{mwconn})$ and $(N,v^{awconn})$ with $\ell_r=10^2$ and $\ell_s=10$.}
		\label{rank7_BOwen_complete}
	\end{center}
\end{table}